\documentclass[namedreferences]{solarphysics}
\usepackage[optionalrh]{spr-sola-addons} 
\usepackage{textcomp}
\usepackage[mediumspace,mediumqspace,Grey,squaren]{SIunits}

\def\nswe#1#2#3{#1\,$#2^\circ\,#3'$}

\usepackage{epsfig}          
\usepackage{graphicx,color}        
\usepackage{amssymb} 
\usepackage{color}           
\usepackage{url}
\usepackage{setspace}
\usepackage{rotating}
\usepackage{float}
\usepackage{textcomp}
\usepackage[breaklinks=true,colorlinks=true,linkcolor=blue, urlcolor=blue,citecolor=blue]{hyperref}
\begin{document}

\begin{article}

\begin{opening}

\title{Detection of high frequency oscillations and damping from multi-slit spectroscopic observations of the corona}

\author{T. ~\surname{Samanta}$^1$\sep
 J.  ~\surname{Singh}$^1$ \sep
 G.  ~\surname{Sindhuja}$^1$ \sep
  D.  ~\surname{Banerjee}$^{1}$ }

\runningauthor{T. Samanta et al.}
\runningtitle{Multi-slit spectroscopic observation of Corona}
   \institute{$^{1}$ Indian Institute of Astrophysics, Bangalore 560034, India
                     e-mail: \url{tsamanta@iiap.res.in}
   }


\date{Accepted: 12 November 2015, Received: 21 January 2015}

\begin{abstract}
During the total solar eclipse of 11 July 2010, multi-slit spectroscopic observations of the solar corona were performed from Easter Island, Chile. 
To search for ``high-frequency waves'', observations were taken at a high cadence in the green line at 5303~\r{A} due to [Fe~{\sc xiv}] and the red line at 6374~\r{A} due to [Fe~{\sc x}]. 
The data are analyzed to study 
the periodic variations in the intensity, Doppler velocity and line width 
using wavelet analysis. The data with high spectral and temporal resolution  enabled us to study the rapid dynamical changes within coronal structures.
We find that at certain locations each parameter shows significant oscillation 
with periods ranging from 6 - 25~s.
For the first time, we could detect damping of ``high-frequency oscillations''  with periods of the order of 10~s. 
If the observed damped oscillations are due to magnetohydrodynamic (MHD) waves then they can contribute significantly in the heating of the corona.  
From a statistical study we try to characterize the nature of the observed oscillations while looking at the distribution of power in different line parameters.
 
\end{abstract}

\end{opening}


\section{Introduction}
To explain coronal heating, existing theories can be broadly grouped into two categories.
One demands a large number of magnetic reconnections. The other theory argues that the heating is dominated
by the damping of magnetohydrodynamic (MHD) waves. It
is now recognized that the solar atmosphere is highly structured in the presence of magnetic fields and 
it is likely that different heating mechanisms may operate in different solar atmospheric structures. 
Observational tests of a specific
heating mechanism may be difficult because several mechanisms might operate at the same time.
The relative contributions of different heating mechanisms are not currently known, 
so we must first look for their signatures in the data, acknowledging that signatures of more than one may be present.
Refer recent reviews by \citet{2000SoPh..193..139R}, \citet{2005LRSP....2....3N}, \citet{2007SoPh..246....3B} and \citet{2009SSRv..149..229T}.


Since the first detection of coronal oscillations by the \citet{1959ApJ...130..215B} and their subsequent
confirmation by
\citet{1977SoPh...51..121T}, there have been a number
of observational pieces of evidence presented for verifying the widespread
existence of oscillations in the solar atmosphere. Using spectroscopic observation with a 40~cm coronagraph, 
\citet{1983A&A...120..185K} found velocity oscillations with periods of 300, 80, and 43~s but no intensity oscillation in the green line (5303~\r{A}). 
\citet{1994scs..conf..487R} reported intensity oscillations in the green line ranging from 5~s to 5~min, 
which they proposed could result from the existence of waves or small-scale dynamic events like nano-flares. 
In the past, a number of researchers  studied the
``high-frequency wave'' properties in the corona by taking  images in the continuum, green and red (6374~\r{A}) 
emission lines during total eclipses \citep{1997SoPh..170..235S,2002SoPh..207..241P,2009SoPh..260..125S}. 
\citet{1999SoPh..188...89C}  detected intensity oscillations with frequencies in the range of 10 to 200~mHz during the 1998 total solar eclipse, while
\citet{2002SoPh..209..265S} used spectroscopic data to detect Doppler velocity oscillation in the range of 1 to 3~mHz and 5 to 7~mHz in the localized regions, 
and they interpreted these variations due to  propagating waves rather than standing waves.  \citet{2002SoPh..207..241P}  reported
frequencies in the range of 0.75 to 1.0~Hz. 
\citet{1997SoPh..170..235S} found variations in the continuum intensity in 6 frequency components with periods 56.5, 19.5, 13.5, 8.0, 6.1, and 5.3~s.
Shorter periodicities are also observed in the radio-band and in X-rays, particularly
in the range of 0.5 to 10~s \citep{1987SoPh..111..113A,2003ApJ...598.1375A}. 
From space-based observation with EIS on Hinode, \citet{2009A&A...494..355O}  found oscillations over a broad range of frequencies (2 - 154~mHz) throughout an active region corona.
They also noticed that the higher frequency
oscillations having frequency greater than 8~mHz, occur preferentially at the edges of bright loops.
More recently, from a  rocket experiment (Hi-C) which operated for several minutes,  \citet{2013A&A...553L..10M}  reported detection  of transverse waves with period 50 to 200~s.

Using the \textit{Solar Eclipse Coronal Imaging System} (SECIS)
instrument \citep{2000SoPh..193..259P} during the total solar eclipse in 1999,
\citet{2001MNRAS.326..428W,2002MNRAS.336..747W} and \citet{2003A&A...406..709K} reported the presence of
propagating fast magnetoacoustic modes in coronal loops
dominated by 6~s intensity oscillation. 
From the same instrument, \citet{2010SoPh..267..305R}  found periodic
fluctuations with periods in the range 0.1 to 17~s. 
Observational detection of these short period waves using 
SECIS instrument complements the theoretical work by \citet{2003A&A...409..325C}. 
In the numerical work,
\citet{1994ApJ...435..482P,1994ApJ...435..502P} explored the processes of coronal heating by 
damping of the slow and fast mode ``high-frequency MHD waves''. From simulation they  concluded that MHD waves can deposit enough energy for heating under certain coronal conditions {\it e.g} the slow mode
waves with periods less than 300 s in the quiet regions and 100 s
in active regions and fast mode waves with periods less than
75~s in the quiet regions and 1~s in active regions can damp sufficiently fast to provide enough energy for balancing radiative
losses. A year later, \citet{1995SoPh..157..103L} showed that 
the Alfv\'enic-type waves with periods of a few seconds (2-10~s) only dissipate in weak magnetic fields ($ < 15$~G). 
In another paper, \citet{1995SoPh..161..269L} showed that acoustic-type waves can also dissipate if they have periods ranging from tens to hundreds of seconds (15-225~s). 
They achieved this range by varying plasma $\beta$, ratio between gas pressure and magnetic pressure, in their model from 0 to 1, 
which is primarily dependent on the density, temperature and magnetic field strength.

Several studies have been
carried out during solar eclipses for the detection of ``high-frequency coronal waves''
using the visible emission lines, 
but their origin remains elusive. 
We still need to understand if they are present preferentially at specific times and locations, or are ever-present.
Thus, it is important to study their temporal, spatial behaviour and more importantly their damping. 
As the damping of these oscillations can only provide the requisite heating. 
Note that space-based EUV telescopes have typical cadence of 10 s or longer,
which makes it difficult to detect oscillations with periods less than 30 s. 
High-cadence observation of the corona can be achieved during total solar eclipses. The observations during total solar eclipses have the advantage
that coronal emission line profiles are free from photospheric light scattered by the sky and provide ideal opportunities to study these variations.
During the total solar eclipse on 11 July 2010, we have performed a spectroscopic observation of the corona with high cadence to study the ``high-frequency oscillations''.
The experimental set up was similar to the experiment during 2009 solar eclipse as described in \citet{2011SoPh..270..213S}.
Instead of a single slit, this
time we have used a multi-slit with faster cadence to better understand ``high-frequency wave'' properties in the corona. 

\section{Experimental set up and Observations}
\begin{figure}
\centering
\includegraphics[angle=0,width=11.5cm]{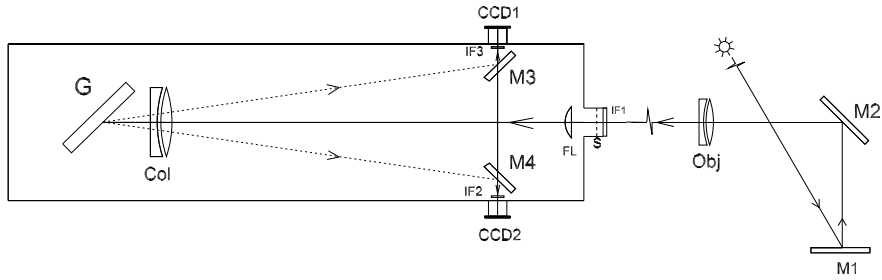}
\caption{Schematic diagram of the optical layout used to obtain spectra around the red and green emission lines of the solar corona. 
M1 and M2 = Flat mirrors of the coelostat system. Obj = Objective lens of 10~cm diameter and of 100~cm focal length. 
IF1 =  Interference filter with transmission in wavelength between 5000-7000~\r{A},  
S = 4 slits separated by 5~mm each, FL = Field lens to focus objective on the collimator, 
Col = Collimator as well as camera lens of the spectrograph, G = Grating with 600 lines mm$^{-1}$ blazed at \unit{2.2}{\micro\meter}, 
M3 and M4 = Flat mirrors to divert the spectral beams, IF2 and IF3 = Narrow band interference filters with FWHM 
of 4~\r{A}, CCD1 And CCD2 = Detectors to record the spectra.}
\label{fig:setup} 
\end{figure}

High resolution spectroscopic observations of the corona in the green 
emission line [Fe~{\sc xiv}] at 5303~\r{A} and the red line  [Fe~{\sc x}] at 6374~\r{A} were carried out
during the total solar eclipse on 11 July 2010 at Easter
Island, Chile, at latitude \nswe{S}{27}{09} and longitude \nswe{W}{109}{26}.
A schematic diagram of the experimental setup is shown in Figure \ref{fig:setup}. 
A two mirror (M1 and M2) coelostat system was used to track the Sun and to direct the sunlight continuously to the spectrograph through an objective lens (Obj).
The alignment of coelostat and tracking speed was selected to achieve negligible movement of the image on the slits of the spectrograph. 
Using an enlarged image of the Sun, we found that drift of the image due to small misalignment and tracking speed over a period of 20 minutes was less than 5$''$. 
This implies that drift of the image was less than 1$''$ during the totality phase of the eclipse which is much less than the width of the slits (20.5$''$). 
A objective lens (Obj) of 10~cm diameter and 100~cm focal 
length formed a 9.3~mm size image of the Sun on the four slits (S) of the spectrograph. 
An interference filter (IF1) with a pass band of about 5000--7000~\r{A} was mounted in front
of the slits to block other light due to higher and lower orders.
The ``slit-width'' of each slit was 100 micron (which corresponds to 20.5$''$ in the Sun) and they were separated by 5~mm from the adjacent slit.
This separation were mutually compatible for both the dispersion of the spectra and pass band of narrow band filters.
Each slit with length of 25~mm permits the recording of spectra up to 2.5 solar radii but the detector size limited the spectra up to about 1.7 solar radii. 
A field lens (FL) just behind the slits avoided over spilling of the beam on the collimator (Col).
A grating (G) of 600 lines mm$^{-1}$ blazed at \unit{2.2}{\micro\meter} and a lens (Col) of 140~cm focal length in Littrow mode provided a 
the spectra.
The 3rd order 6374~\r{A} wavelength and 4th order 5303~\r{A} wavelength regions were selected for observations 
as it was easy to focus by the same collimator lens (Col). The final configuration provided
a dispersion of 3.3~\r{A} mm$^{-1}$ and 2.3~\r{A} mm$^{-1}$ around the 3rd order red and 4th order green emission line, respectively. 
We could not mount the CCD detectors (CCD1 and CCD2) directly on the focused spectral region due limitation of space. 
Therefore, we used 75~mm flat mirrors (M3 and M4) to divert the red and green wavelength regions of the spectrum as shown in Figure \ref{fig:setup}
and mounted two CCD detectors. 
Two narrow band interference filter (IF2 and IF3) with pass band of 4~\r{A} 
centred around 6374~\r{A} for the red line and 5303~\r{A} for the green line were installed in front 
of detector (CCD1 and CCD2) to avoid the overlap of spectra due to two adjacent slits.

\begin{figure}
\centering
\includegraphics[angle=90,width=9.5cm]{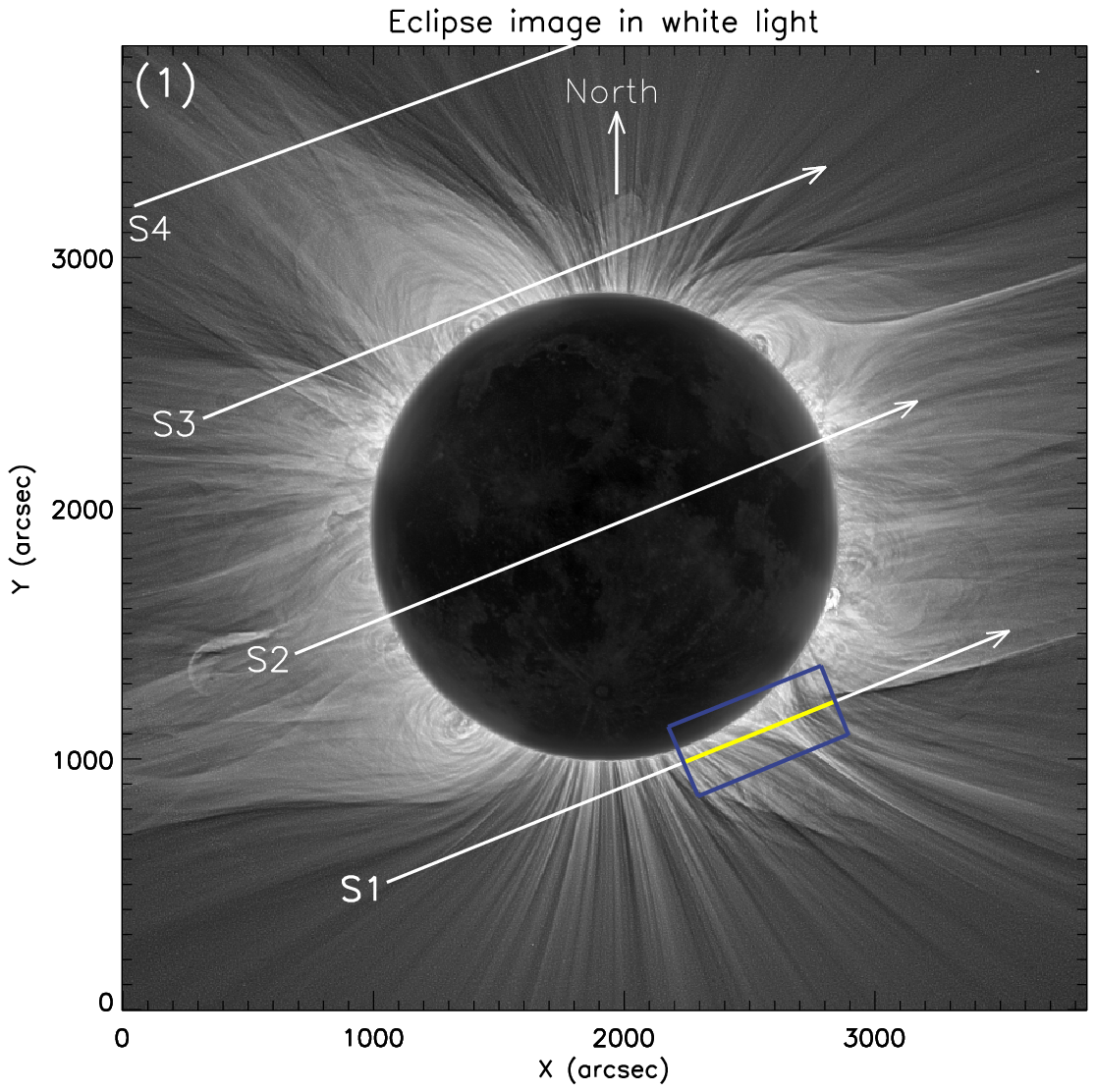}
\includegraphics[angle=90,trim = 0mm 0mm 0mm -2.5mm, clip,width=9.5cm]{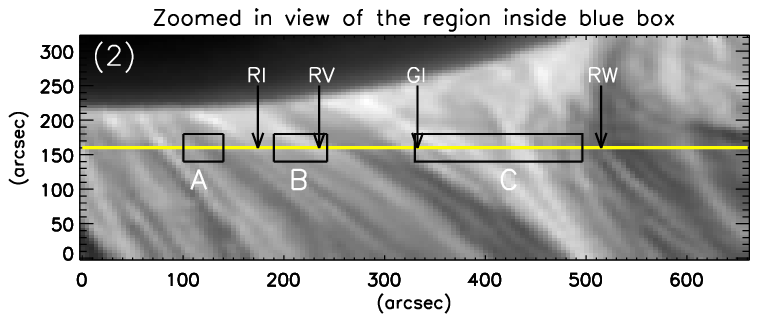}
\caption{(1): White light image of the corona taken during the total eclipse of 11 July 2010. 
This image was obtained by Dr. Miloslav
Druckm\"{u}ller$^1$
form the Tatakoto Atoll, French Polynesia.
Position of four slits (S1-S4) from our spectroscopic experiment setup are shown by the white lines.
The yellow line (on top of the S1) marks the region where the signal is good. This is our region of interest (ROI).
 The arrow indicates ascending pixel numbers along the length of the slit.
The lower panel show a zoomed in view of the blue box as marked in the upper panel. The  RI, RV, RW and GI arrows marked on the slit show the locations where  we performed wavelet analysis and results are shown in Figure
\ref{fig:int_wavelet}, \ref{fig:vel_wavelet}, \ref{fig:wid_wavelet} and \ref{fig:int_wavelet_green} respectively.
A, B and C marks three structures as identified from the intensity space-time plot of the red line (see Figure~\ref{fig:int_xt}).}
\label{fig:slit_posn} 
\end{figure}
A \unit{13}{\micro\meter} pixel size of the CCD  detector was capable of providing a resolution of 0.043~\r{A} 
in the 3rd order of red line but the slit width of \unit{100}{\micro\meter} limited the spectral resolution to 0.33~\r{A}. The pixel resolution along the slit is 2.64$''$ for the red line spectra. 
The EM-CCD camera of ANDOR of 1k x 1k format with 14--bit 
read out at 10~MHz was used in the frame transfer mode for taking the spectra around the red emission line with a cadence of 1.013~s 
(Exposure time of 1~s and 0.013~s for frame transfer). The gains of the EM-CCD camera was set at 200 to enhance the signal to a reasonable level.
The EM (Electron Multiplication) detector magnifies the week signals with some increase in the noise. The net result is that 
it provides possibility to study the week signals with very high temporal resolution.
We had only one CCD camera that could be operated in frame transfer mode but had another detector without EM facility. 
We, therefore, used another CCD camera (ANDOR) of 2k x 2k format with a \unit{13.5}{\micro\meter} pixel$^{-1}$ size for recording the spectra in the green line. 
The read-out speed was less, 2~MHz with 16--bit data. The chip was binned 2 x 2 to decrease the read time by a factor of two  
and region was also reduced to 75 \% to decrease the read-out time further. The binned detector had a pixel resolution of 0.062~\r{A} 
but the slit width restricted the spectral resolution to 0.23~\r{A} in the green channel. Green line spectra have 5.52$''$ pixel resolution along the slit.
The spectra in the green emission line were 
recorded with a cadence of 3.64~s (exposure time of 3~s and read-out time of 0.64~s including shutter operation). 
It may be noted that the estimate of the exposure times in both the cases was obtained by making the observations during the period of 
full moon at night while testing the experimental setup. 
Dark signal was obtained for calibration by closing the slits of the spectrograph and recording data under the same conditions.
The solar disc spectra were also obtained during a period of 
clear sky to convert the observed coronal intensity to the absolute units. The 27~mm size of the detector permitted us to record the spectra up to $\sim$~3 solar radii in the green channel.
The short exposure time could not allow obtaining spectra with the the fourth slit because of decrease in 
the emission line intensity, being too far above the solar limb. 
\begin{figure}
\centering
\includegraphics[angle=90,width=6.05cm]{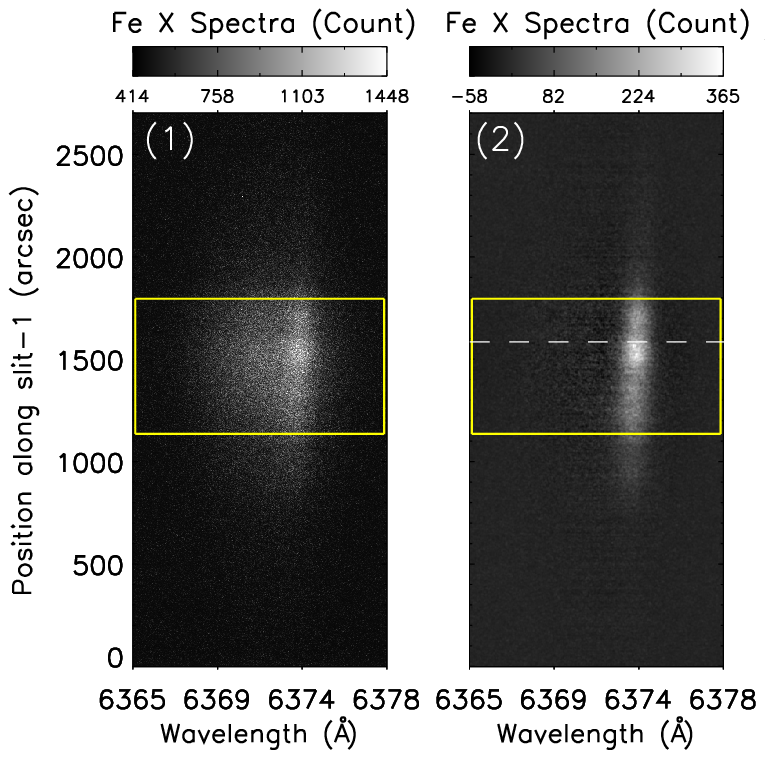}\includegraphics[angle=90,width=6.15cm]{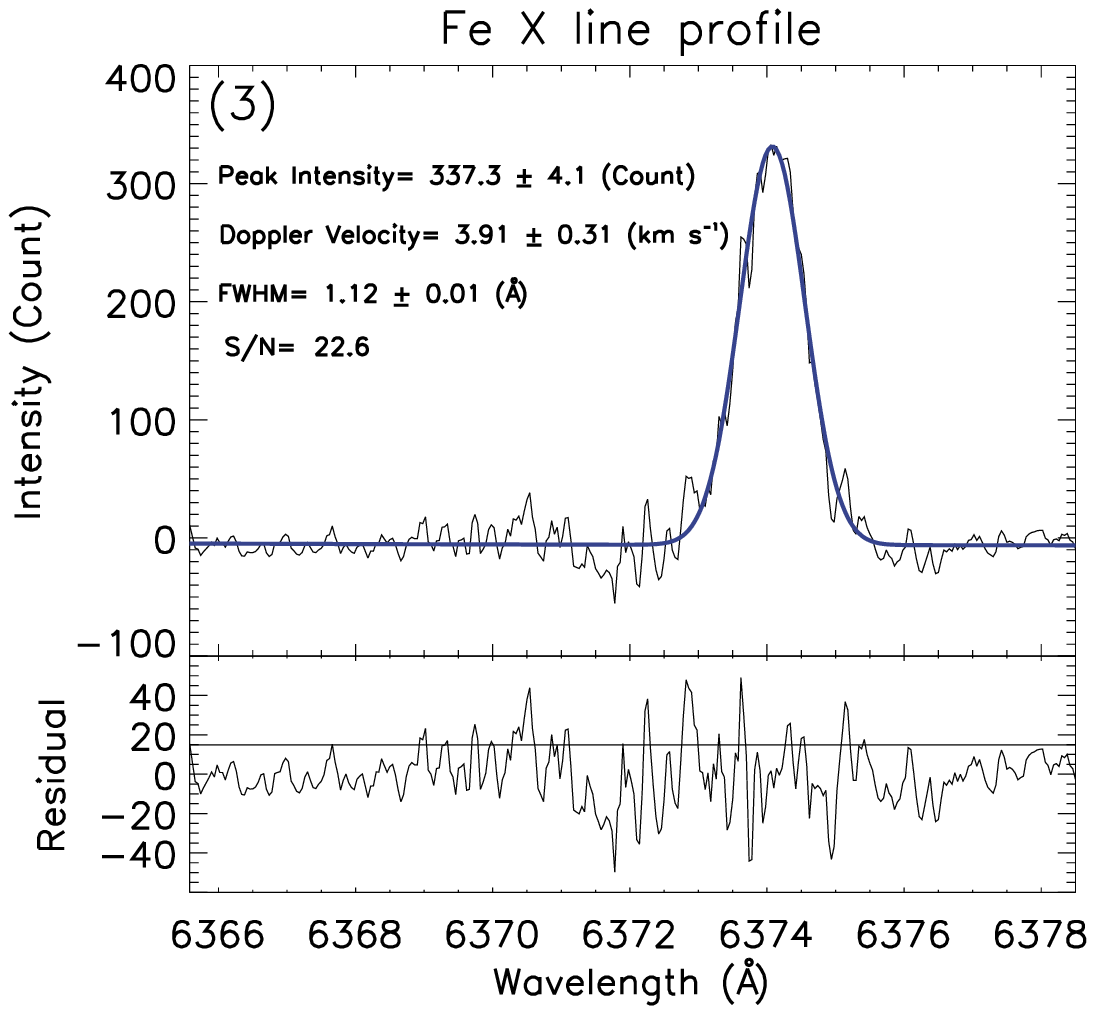}
\caption{Panel (1) shows the raw red line spectrum taken with the first slit (S1) recorded during the total phase of the solar eclipse.
(2) shows the spectrum after accounting for dark current and the transmission curve of the narrow band interference filter. 
The line profile along the white dashed line is shown in the right panel. The yellow rectangular box shows our ROI in the blue box shown in Figure~\ref{fig:slit_posn}.
(3) \textit{Top panel:} shows the red line profile. The Gaussian fit to the profile is drawn in blue. 
Extracted line parameters from the line profile fitting are printed.
\textit{Bottom panel:} shows the residual between the fitted curve and 
the original line profile. Horizontal line represents the absolute standard deviation of the residual.}
\label{fig:a}
\end{figure}
At the location of the third slit, it appears that intensity of the emission corona was not sufficient to make a distinct impression 
of  the emission component over the continuum background. Finally we obtained spectra with two slits as shown in Figure~\ref{fig:slit_posn}.
The slit locations have been marked by the white lines on the broad band image of the solar corona as was obtained by Dr. Miloslav
Druckm\"{u}ller\footnote{\url{http://www.zam.fme.vutbr.cz/~druck/eclipse/Ecl2010t/Tse2010t_1000mm_1/0-info.htm}} form the Tatakoto Atoll, French Polynesia
\citep{2012SoPh..278..187V, 2011ApJ...734..120H}. 
The yellow portion marked on the first slit (S1) indicates the portion of the solar corona where the emission 
spectra was strong and we could analyze the data reliably.

\section{Data Reduction}

First, the coronal spectra and all the disk spectra were corrected 
for the dark current by subtracting the respective dark signal. 
After that, the spectra due to the four slits were divided into four spectral windows. 
\begin{figure}
\centering
\includegraphics[angle=90,width=10.5cm]{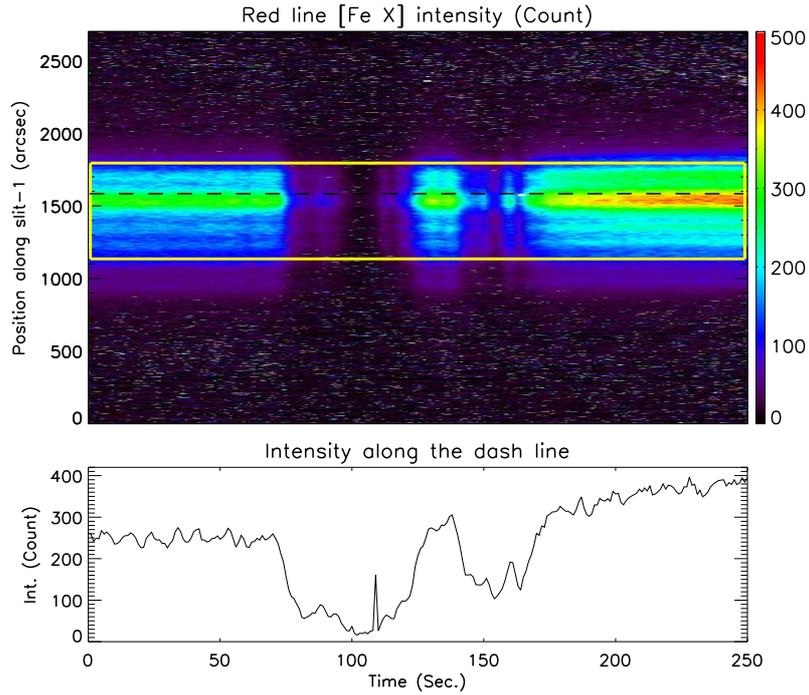}
\caption{First panel shows the temporal evolution of the red line intensity at each pixel along the slit-1.
The origin of this plot is used as reference for selecting the time interval and choosing the spatial locations for all the further analysis.
The bottom panel shows the intensity variations along the horizontal black dashed line. The sudden drop of intensity during the middle part of totality is due to 
passing clouds which reduces the intensity. Yellow rectangular box shows the ROI as mentioned in Figure~\ref{fig:slit_posn}.}
\label{fig:b} 
\end{figure}
The raw spectrum window due to the first slit (S1) is shown in Figure~\ref{fig:a}-(1).
The use of narrow band filters (IF2 and IF3) in front of CCD cameras modified the continuum part of the spectra. 
We have determined the transmission curve of each filter using the solar spectra at different parts of the corona inside our field of view (FOV). 
There was a small variation in the transmission curve of the filter as a function of the FOV,
but it was possible to use the mean transmission curve of the filter for the observed solar corona.  
The intensity of the observed profile in the two wings of the spectrum away from the emission line 
was found out for each location on the slit and then the transmission curve of the 
filter was normalized to match the observed intensity of the profile at these wavelengths. 
Then, the derived transmission curve was applied to the observed profile at each location on 
the slit to make the spectra free from the effect of transmission curve of the filter. 
The corrected spectrum after accounting for the transmission curve of the narrow band interference filter is shown in Figure~\ref{fig:a}-(2).

Gaussian fits were applied to the line profiles at each pixel to derive the peak intensity, 
line-width (FWHM) and Doppler shift of the emission line's centroid with respect to the reference wavelength.
We have averaged all the Gaussian peak positions to determine the reference wavelength of the lines.
The averaged line centers are taken as 6374.4~\r{A} and 5302.8~\r{A} for the red and green line, respectively.
The upper panel of Figure \ref{fig:a}-(3) shows an example of 
red line profile after the correction for the dark current and the transmission curve of the narrow band interference filter.
The blue curve represents a Gaussian fit to the line profile. All the
extracted line parameters and their fitting errors (1$\sigma$) in the measurements are printed in the same panel. The signal to noise ratio (SNR) is also given.
After extracting the line parameters for each of the locations on the slit-1 for
each time frame, the temporal evolution along the slit is shown in the upper panel of Figure \ref{fig:b}.
The sky
conditions were good but sudden drops of intensity 
during the
middle part of totality indicates a thin passing cloud which reduced the intensity at that time.
The bottom panel shows intensity variation at a typical location during our observation.
Low signal during the passing of the cloud did
not permit a good fit to the emission line. We, therefore obtained good data in two time intervals, both of 70 s duration (the first 70 s of totality and the last 70 s of the totality). 
We have separately analyzed these two sets of data only at the region inside the ROI. The SNR in this two intervals varies from 8 to 29.

\section{Results}
\subsection{Detection of oscillations}
\begin{figure}
\centering
\includegraphics[angle=90,width=9.5cm]{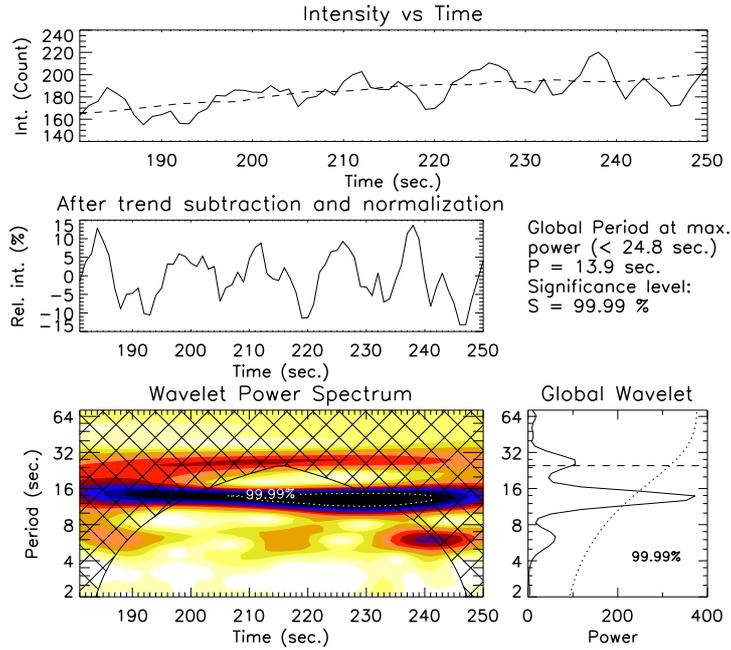}
\caption{A typical example of wavelet results corresponding to RI  location  (marked in Figure~\ref{fig:slit_posn}) and in the red line.
In the top panel the variation of intensity with time is represented by a solid line. The dashed line represents 
a background trend. The middle panel shows the normalized intensity variation.
The bottom left panel shows the wavelet power spectrum of the normalized time series. Overplotted cross-hatched regions
above the wavelet power spectrum is the cone of influence (COI).
The location of power above 99.99 \% significance level
is indicated by the region overplotted
with dotted white line contour. Note that darker colour represents higher power. 
The bottom right panel shows the global wavelet power.
The maximum measurable period is 24.8 s (arises due to COI) which is shown by a horizontal
dashed line. The dotted line
above the global wavelet power plot shows the significance level of 99.99 \%. 
The significant periods as measured from the
global wavelet power is printed on the top of the global wavelet power plot. 
Note that the power is unitless as the wavelet transformation is applied to normalized time series.
}
\label{fig:int_wavelet}
\end{figure}
\begin{figure}
\centering
\includegraphics[angle=90,width=9.5cm]{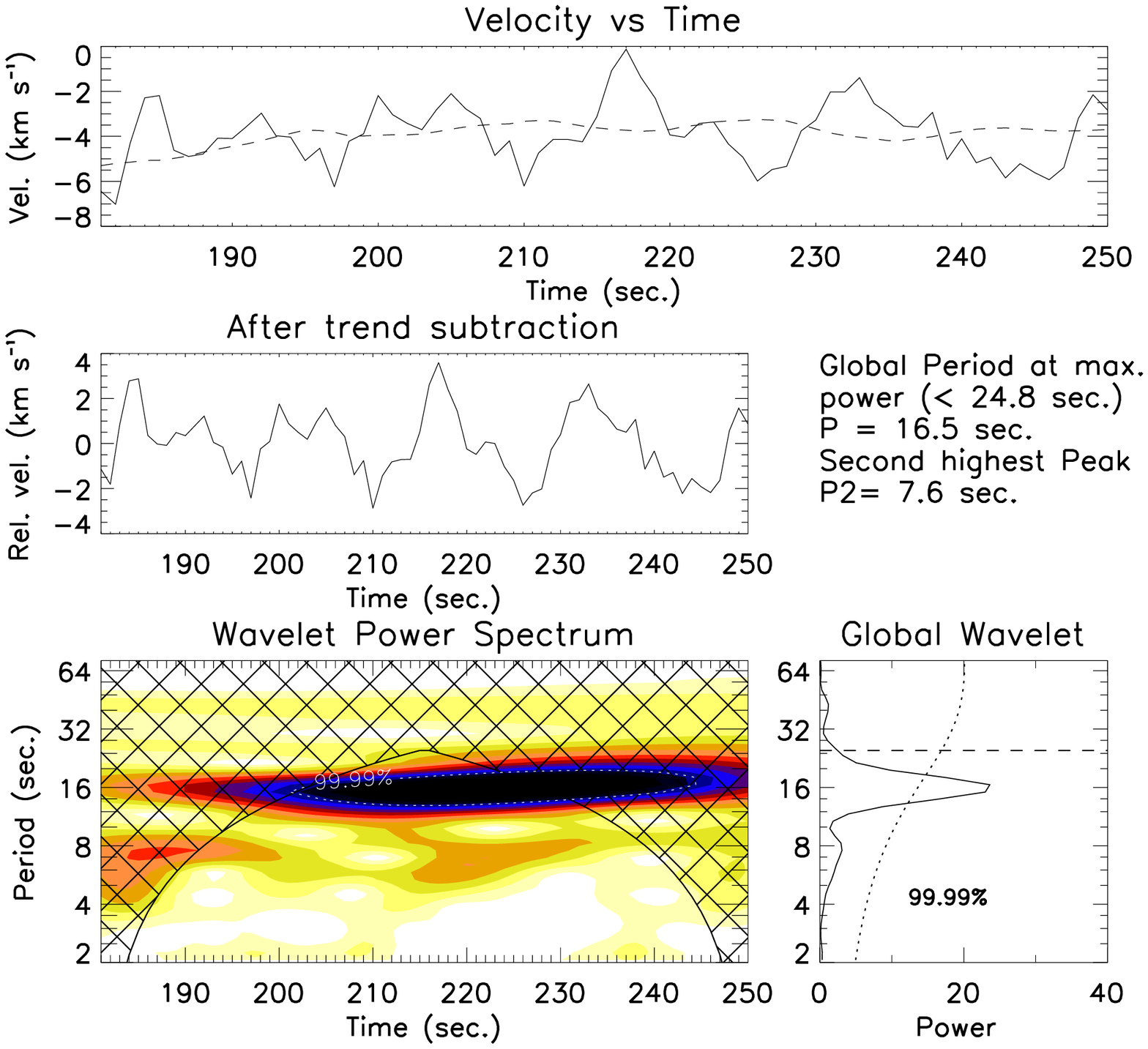}
\caption{A typical example of the Doppler velocity variations of the red line and 
their wavelet analysis for the location RV indicated in Figure~\ref{fig:slit_posn}. 
The panels are as in Figure~\ref{fig:int_wavelet}.}
\label{fig:vel_wavelet}
\end{figure}

We have applied wavelet technique \citep{1998BAMS...79...61T} for time series analysis at each location inside the ROI
and for each line parameter (peak intensity, FWHM and Doppler velocity).
We have used a Morlet function, a complex sine wave modulated
by a Gaussian, for convolution with the time series in the wavelet transform.
Figure \ref{fig:int_wavelet} shows a typical example of the result from wavelet analysis. 
In the top panel, the variation of intensity with time (I) and the  background trend (I$_{bg}$), are shown.
The middle panel shows relative intensity variations ($I_{R}$) with respect to the background trend  ($I_{R}=[I-I_{bg}]*{I_{bg}}^{-1}*100 $~\%).
The cross-hatched regions in the wavelet power spectrum called cone of influence (COI), is 
the region where the power is not reliable and it arises due to the finite length of time series. 
The bottom right panel shows the global wavelet power which is the time averaged wavelet power at each
period scale.
\begin{figure}
\centering
\includegraphics[angle=90,width=9.5cm]{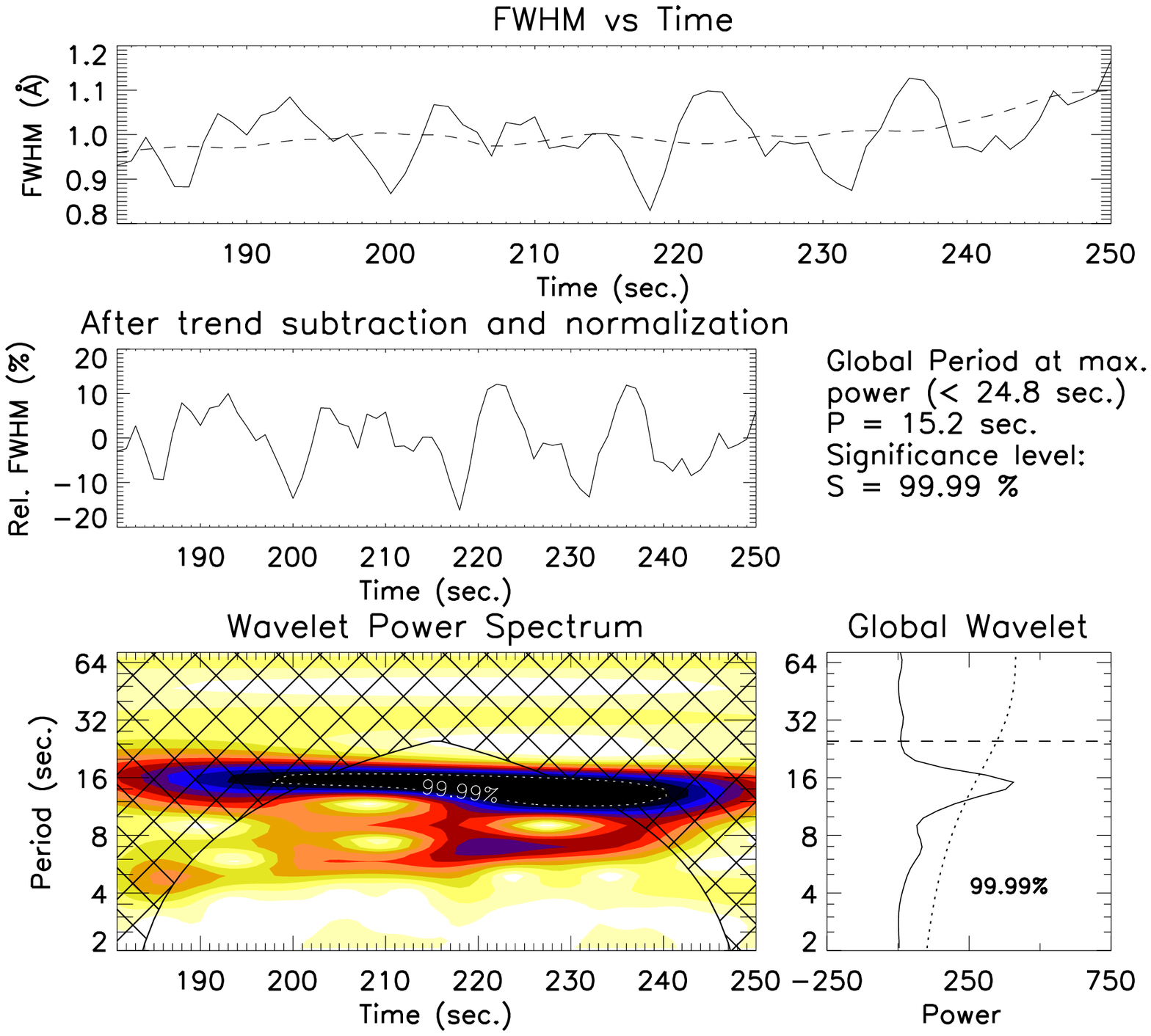}
\caption{A typical example of the FWHM variations of the red line and their wavelet analysis for the location RW indicated in Figure~\ref{fig:slit_posn}.The panels are as in Figure~\ref{fig:int_wavelet}.}
\label{fig:wid_wavelet}
\end{figure}
\begin{figure}
\centering
\includegraphics[angle=90,width=9.5cm]{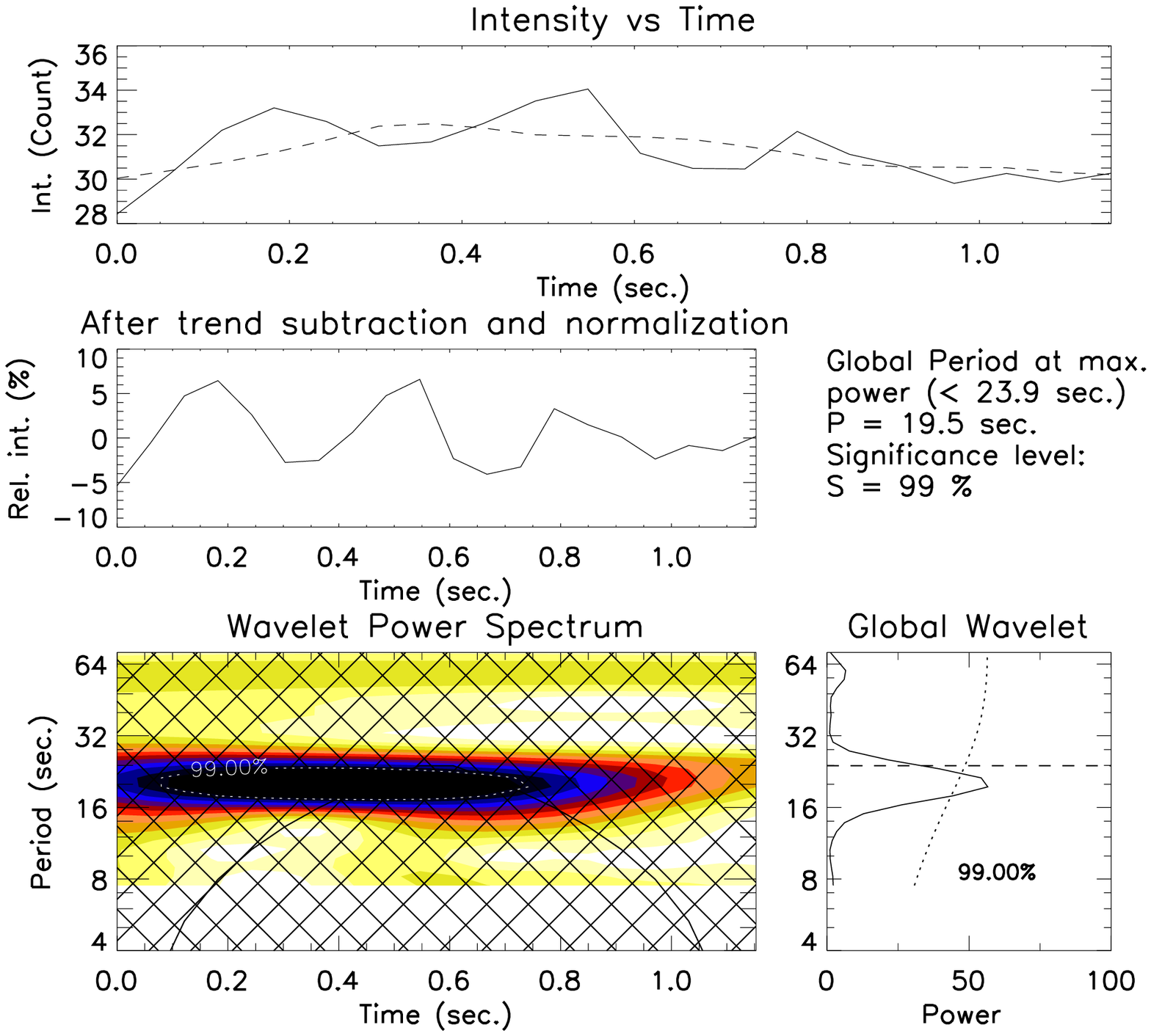}
\caption{A representative example of intensity variations as recorded from the green line spectra and their wavelet analysis 
results for the location GI indicated in Figure~\ref{fig:slit_posn}. The panels are as in Figure~\ref{fig:int_wavelet}.}
\label{fig:int_wavelet_green}
\end{figure}
The dotted line above the global wavelet power plot shows the significance level of 99.99 \% 
calculated by assuming a white noise \citep{1998BAMS...79...61T}. 
The white noise is a random distribution about the mean of the original time series, which has a flat Fourier spectrum. 
If a peak in the wavelet power spectrum is significantly above this background white noise spectrum, 
then it is assumed to be a real feature with a certain percentage of confidence (see \citet{1998BAMS...79...61T} for details).
Because of the total duration of the time series being 70 s, the COI effect restricts us to measure the significant period only up to 25 s.
Keeping this in view, we have subtracted a 30-point running average  (the background trend)
from the original time series to
suppress the variations above 30 s from the time series.
The final result of the analysis shows that a oscillation with 14 s periodicity is present in the intensity variation throughout the observing period of 70 s.
Similarly, Figures \ref{fig:vel_wavelet} and \ref{fig:wid_wavelet} show that
16 s and 15 s periodicities are present in the Doppler velocity and FWHM variations  respectively.

A typical example of the wavelet analysis for the green line intensity oscillation is shown in Figure \ref{fig:int_wavelet_green}. 
The green line spectra were taken with a different camera with a lower cadence of 3.64 s. 
As the data points for the green line were fewer than that for the red line whose cadence was 1.013 s, 
the confidence level for detection is not as good as  the red line. We have studied oscillation in all the green line parameters. 
The result shows that the oscillations are detectable for all the line parameters
with periods ranging from 10-25 s, but we have used a lower significance threshold of 99 \% for  detection.
Hereafter, we have concentrated only on red line data which has more data 
points and provides reliable confidence level. 
In this section we have shown that we have detected significant oscillations at isolated points
but for a proper diagnostics of the wave modes which could be responsible for these oscillations we need a more statistical 
approach which we address in the following subsection. 
\label{sec:1}

\subsection{Statistical behavior of the oscillations}
\begin{figure}
\centering
\includegraphics[angle=90,width=12.0cm]{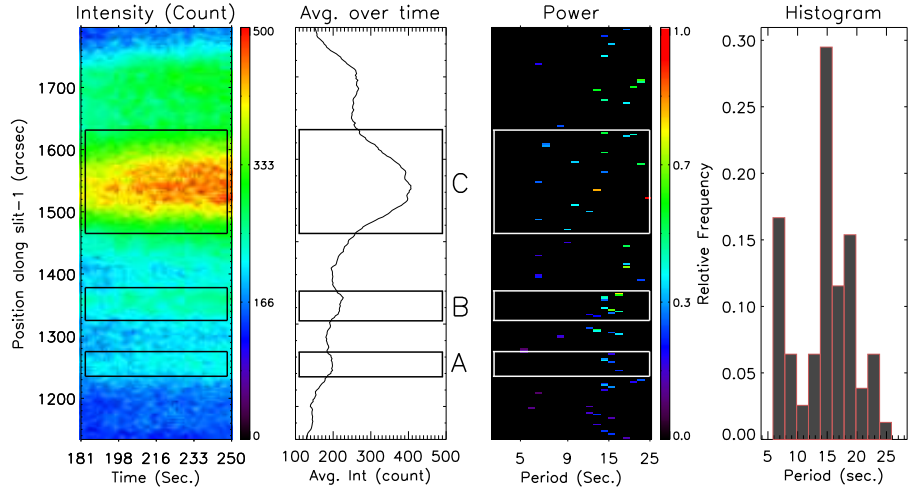}
\caption{\textit{Left to right:} The first panel shows the temporal evolution of the red line intensity along slit 1. 
This is similar to Figure~\ref{fig:b}, but for a region inside the ROI and for the last 70~s time interval of totality. 
The second panel shows the time-averaged intensity variation along the slit. 
The third panel shows the dominant periods of oscillation above 99.99 \% significance level. 
Color indicates the amplitude of the power (normalized).
The last panel shows a histogram of the distribution of the significant periods.}
\label{fig:int_xt}
\end{figure}
\begin{figure}
\centering
\includegraphics[angle=90,width=12.0cm]{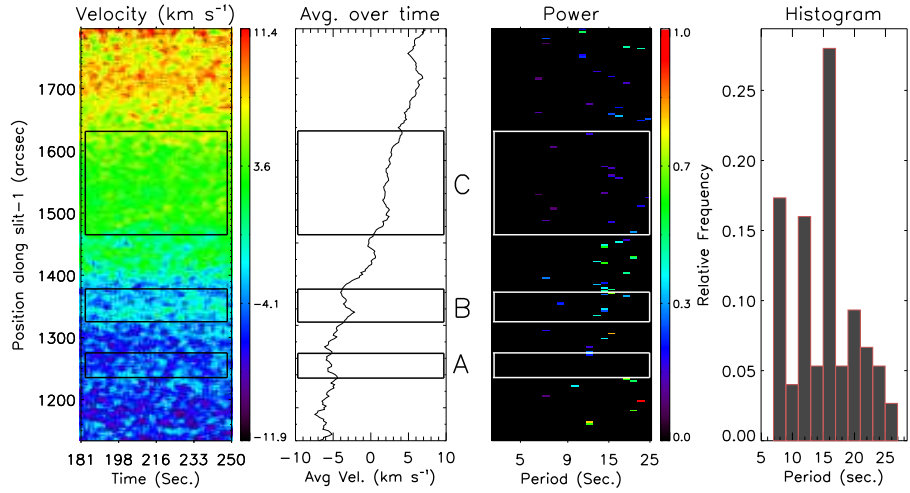}
\caption{Similar to Figure~\ref{fig:int_xt}, it shows the temporal variations of Doppler velocity along the slit and its statistical behavior of the oscillation.}
\label{fig:vel_xt}
\end{figure}
\begin{figure}
\centering
\includegraphics[angle=90,width=12.0cm]{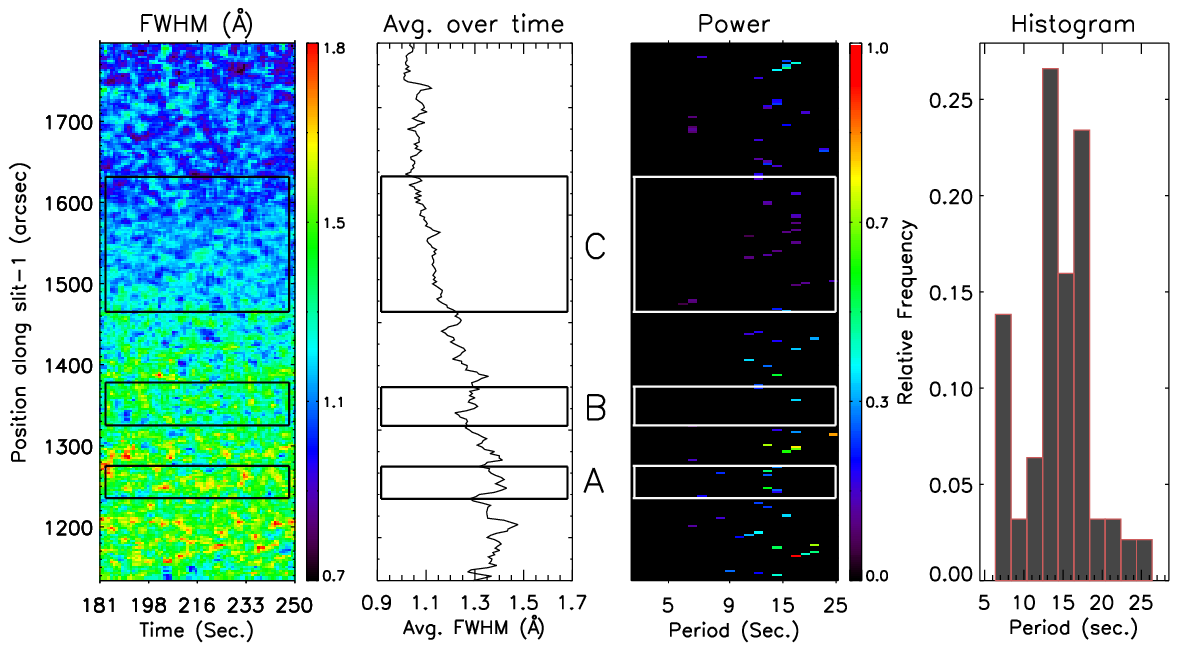}
\caption{Similar to Figure~\ref{fig:int_xt}, it shows the temporal variations of FWHM along the slit and its statistical behavior of the oscillation.}
\label{fig:wid_xt}
\end{figure}

\begin{figure}
\centering
\includegraphics[angle=90,width=12.0cm]{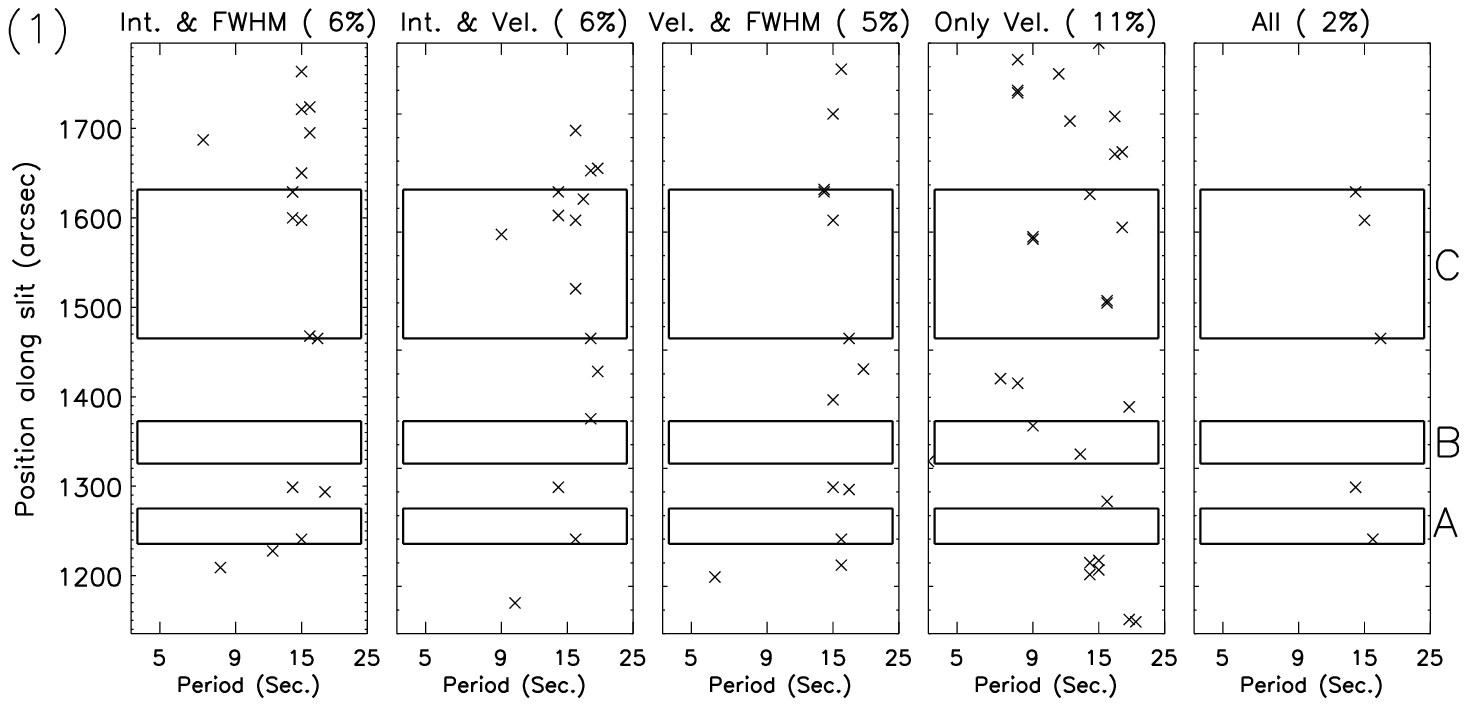}
\includegraphics[angle=90,width=12.0cm]{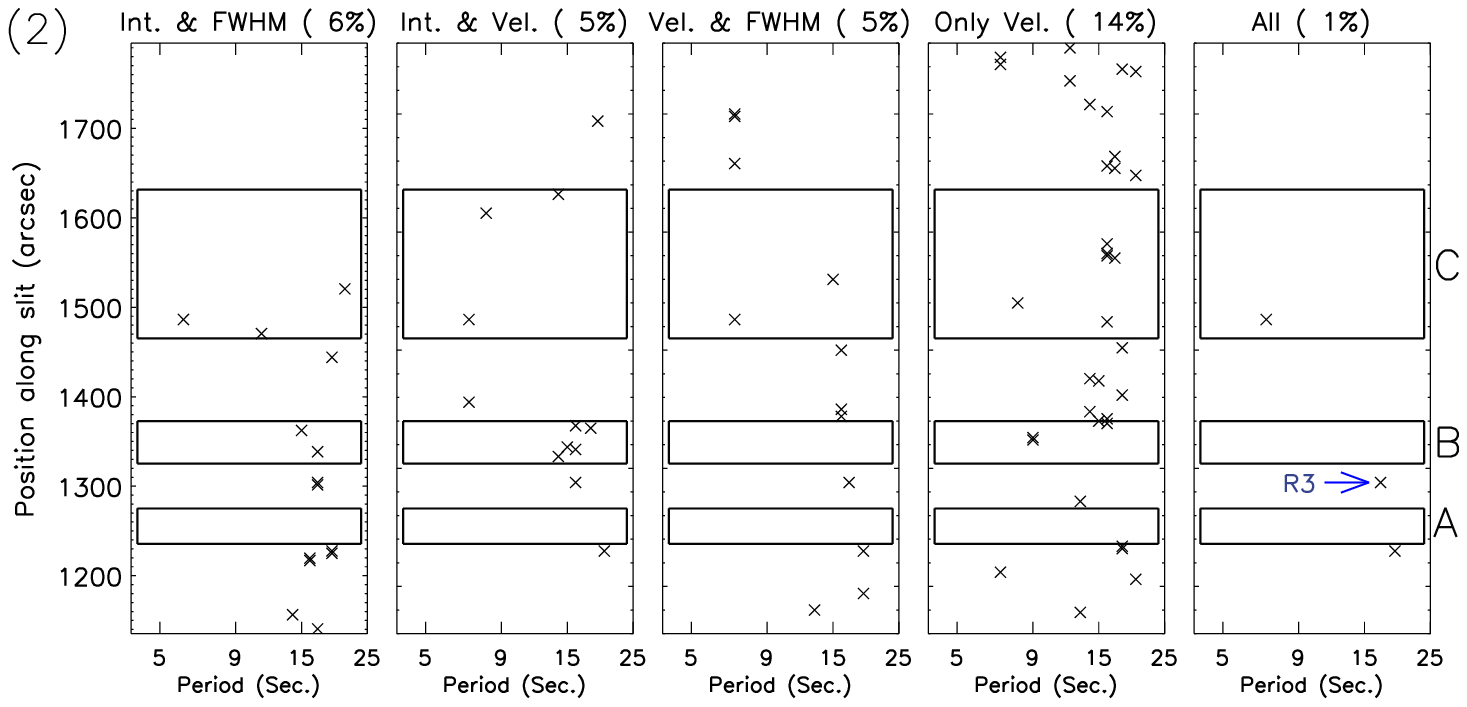}
\caption{The top row (1) corresponds to analysis result for first 70 seconds of totality and the bottom row (2) corresponds to 181 to 250 s of totality. 
From left to right: the locations where both the intensity and width oscillations are present with identical periods, the
second panel showing the locations where
both intensity and Doppler velocity oscillations are present with identical periods, the third panel showing the locations where
both width and Doppler velocity oscillations are present with identical periods, the fourth panel shows the locations where only Doppler velocity is present and the last panel shows
the locations where oscillations are present with identical periods in all the parameters.
The numbers in parenthesis on top of each panel represent the percentage of pixels where we detect significant oscillation.}
\label{fig:int_wid_dop}
\end{figure}

In this subsection, we study the statistical properties of the oscillations and whether
these oscillations are preferentially located within some structures or near the boundaries. 
In this context, we have produced a space-time (XT) map from the red line intensity.
The first panel (from left) of Figure~\ref{fig:int_xt} shows the temporal evolution of intensity along the slit (intensity XT map). 
The time averaged intensity profile along the slit (S1) is shown in the second panel.
Its variations along the slit indicates that the slit crosses some structures which is also clear 
from the context white light eclipse image (Figure~\ref{fig:slit_posn}).
While looking at the  intensity variation along the slit, we have identified three structures A, B and C.
We should point out that  determination of the exact boundary is not possible from our sit and stare observation. 
Note that we do not have simultaneous imaging observation from the same location and hence we can only indicate approximate boundary while 
looking at the  intensity variation  along the slit. 
Furthermore, one should note that, the red line emission profile (variation along the slit) may not exactly match what we see in white light.  
In Figures~\ref{fig:int_xt},~\ref{fig:vel_xt}~\ref{fig:wid_xt}~and~\ref{fig:int_wid_dop} the 
rectangular boxes from bottom to top represent structures A, B and C respectively.  
The structure C is wider and may have more than one overlapping structures.

We have performed wavelet analysis to look for oscillation signatures at each pixel along 
the slit to determine the distribution of the period of oscillations and their power.
First, we have calculated a global wavelet power spectrum at each 
pixel and afterwards the power map has been determined by the pixels
where global power exceeds the significance level. 
The third panel of Figure~\ref{fig:int_xt} shows the power map for the intensity variations.
This figure provides an overview of distribution of power at different locations along with significant periodicities. 
The third panel of Figures~\ref{fig:vel_xt} and \ref{fig:wid_xt} show the spatial distribution of power for Doppler velocity and width oscillations, 
similar to the power distribution for intensity as shown in the third panel of Figure~\ref{fig:int_xt}.
The power map of intensity variations shows that the intensity oscillations have small preference to occur in the thinner structure A and B 
but not in the wider structure C. 
Whereas the FWHM and Doppler velocity power map shows that they have 
a slight tendency to occur close the boundaries of the structures where the intensity gradient is relatively high.
The result is based on a comparison of three coronal structures only. Hence we are not able to conclude further on the statistics.
We have also noticed that the oscillations at periods less than 12 s are barely present in  A and B but more frequent in the extended structure C.
It has been pointed out \citep{2001A&A...368.1095O,2009SoPh..260..125S,2009A&A...494..355O,2010SoPh..267..305R} 
that the  intensity oscillations are significantly prevalent at the edges of bright coronal loops.

To find the distribution of the time periods, we have made histogram plot with a 2 s binning in the time period domain.
The histogram for the intensity, Doppler velocity and width oscillations are shown in the last panels of Figures~\ref{fig:int_xt}, \ref{fig:vel_xt} and \ref{fig:wid_xt} respectively.
It provide an estimate on which periodicity is statistically most prevalent. 
It shows two peaks, one around 14 to 20 s and the other
around 6-8 s. \cite{2001MNRAS.326..428W} reported a peak around 6 s in the intensity oscillation data.


\subsection{Damping signature of the oscillations}

\begin{figure*}
\centering
\includegraphics[angle=90,width=6.1cm]{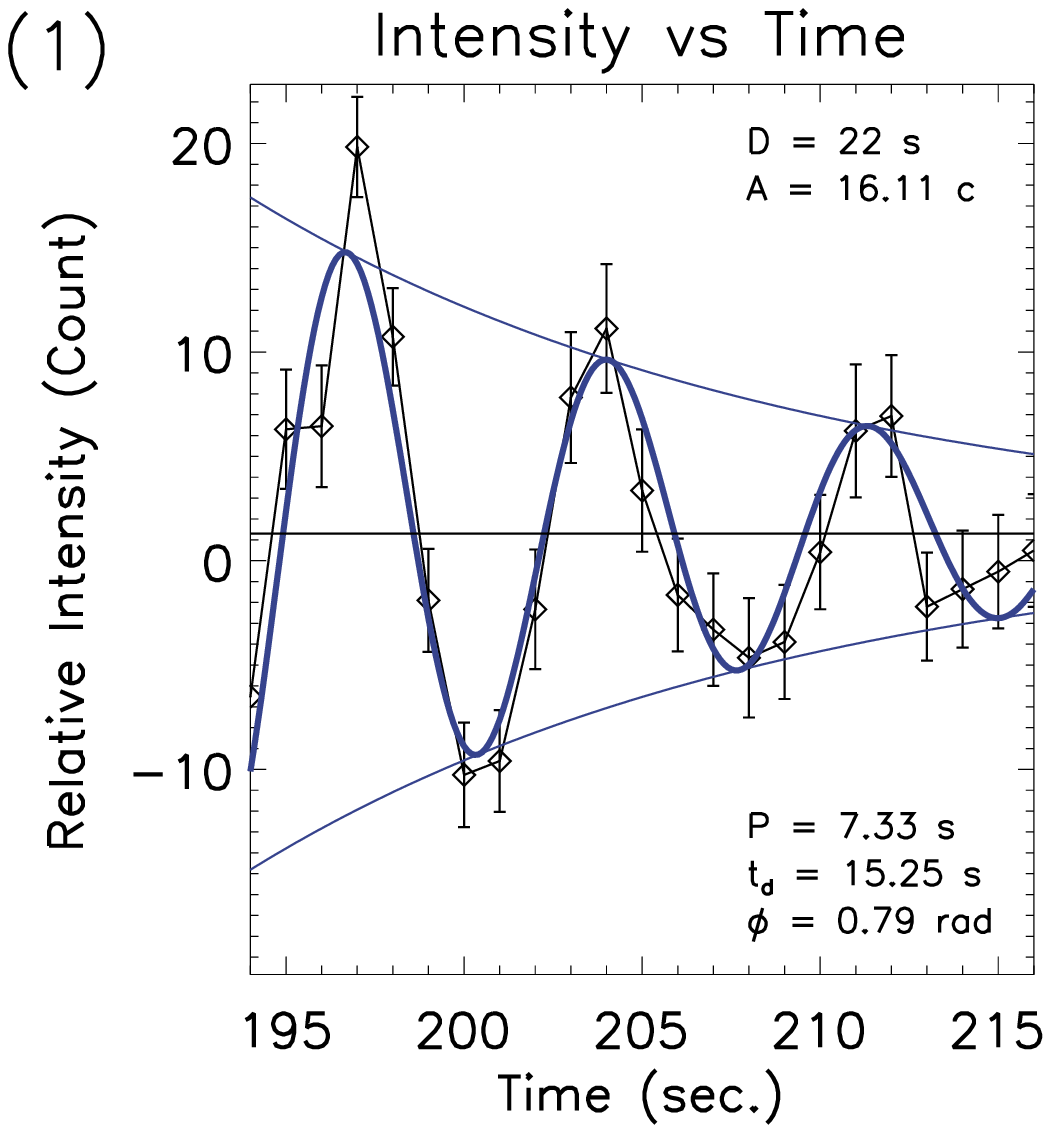}\includegraphics[angle=90,width=6.1cm]{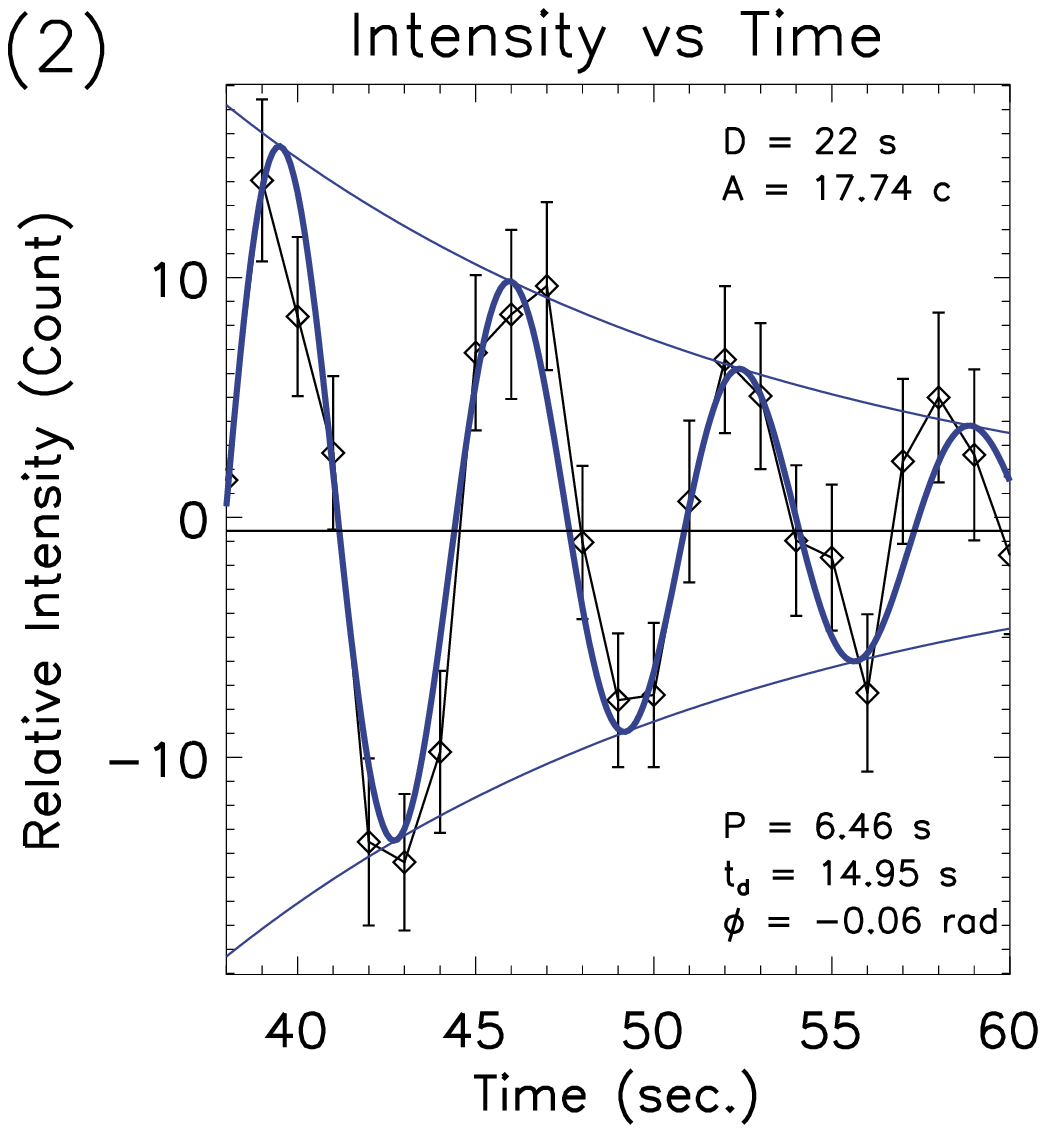}
\includegraphics[angle=90,width=6.1cm]{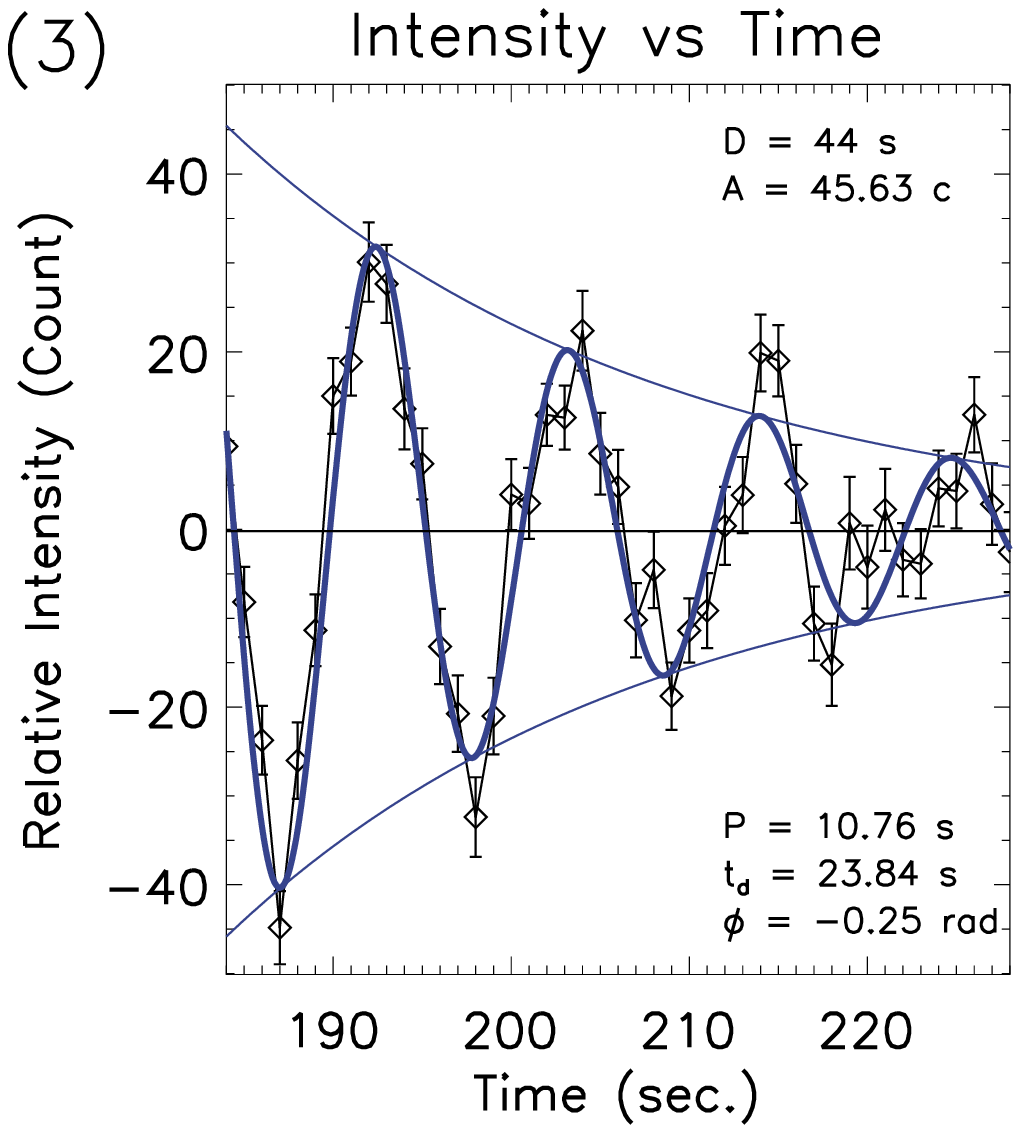}\includegraphics[angle=90,width=6.1cm]{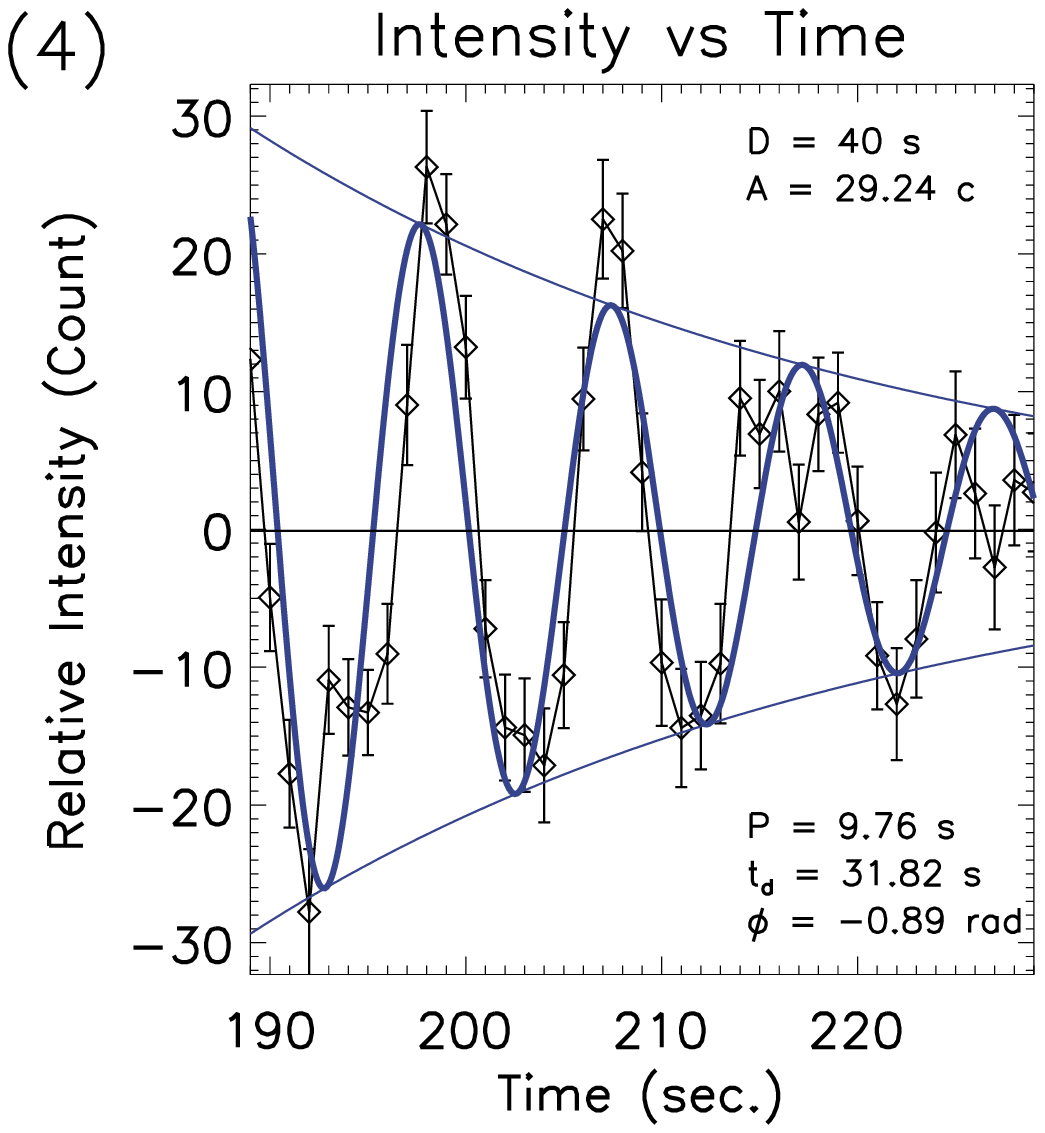}
\caption{Examples of damped intensity oscillations. The location of the occurrence (1-4) are shown by the arrow marks I1, I2, I3 and I4 in Figure~\ref{fig:int_xt_d}. 
The data points (diamonds) are fitted with damped sine functions (see Equation~\ref{eqn1}) represented  by thick blue curves.
The error bars calculated from Gaussian fitting are also shown. The damping
parameters from the fitting are printed in each panel.}
\label{fig:int_damp}
\end{figure*}
\begin{figure*}
\centering
\includegraphics[angle=90,width=6.1cm]{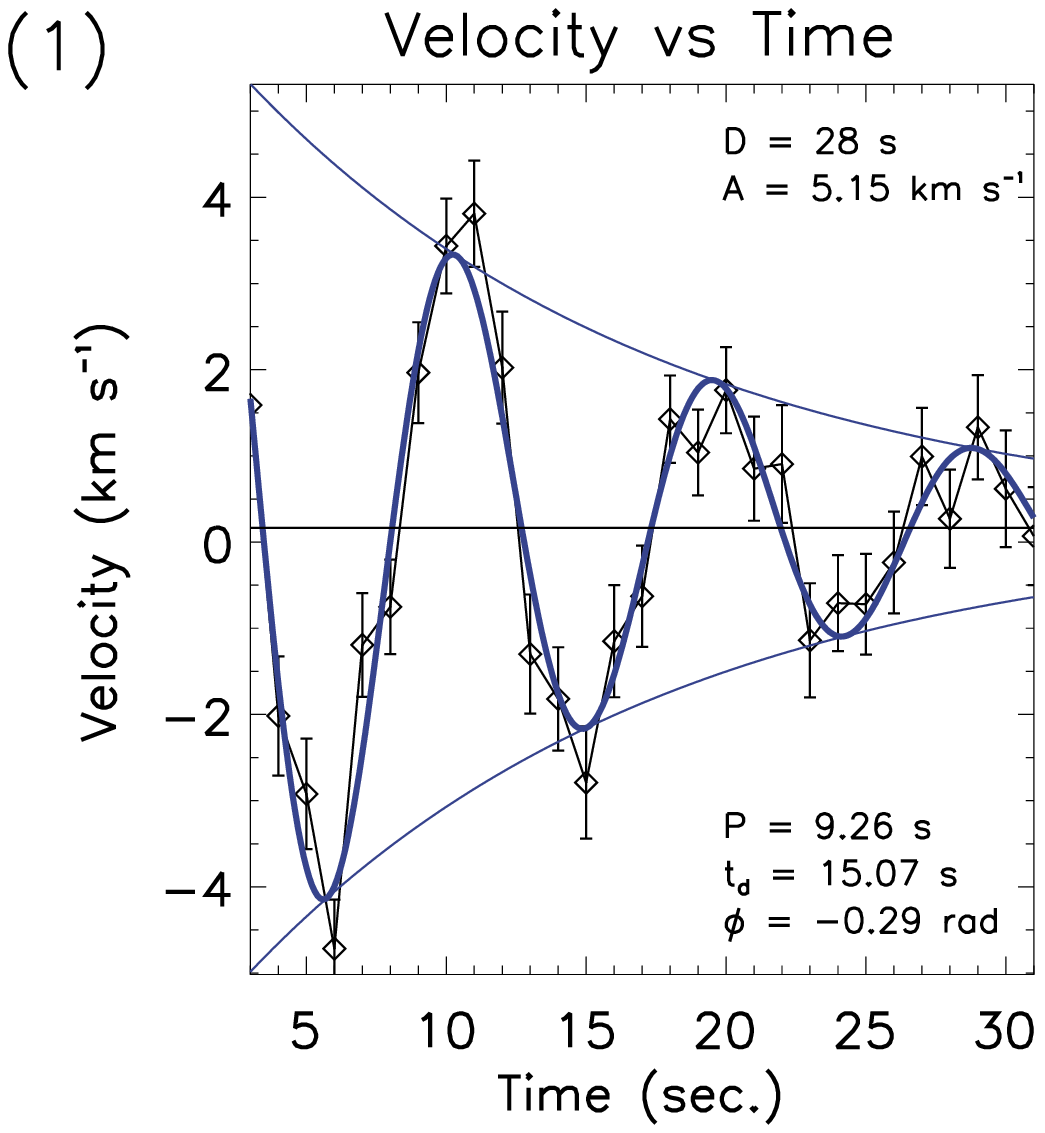}\includegraphics[angle=90,width=6.1cm]{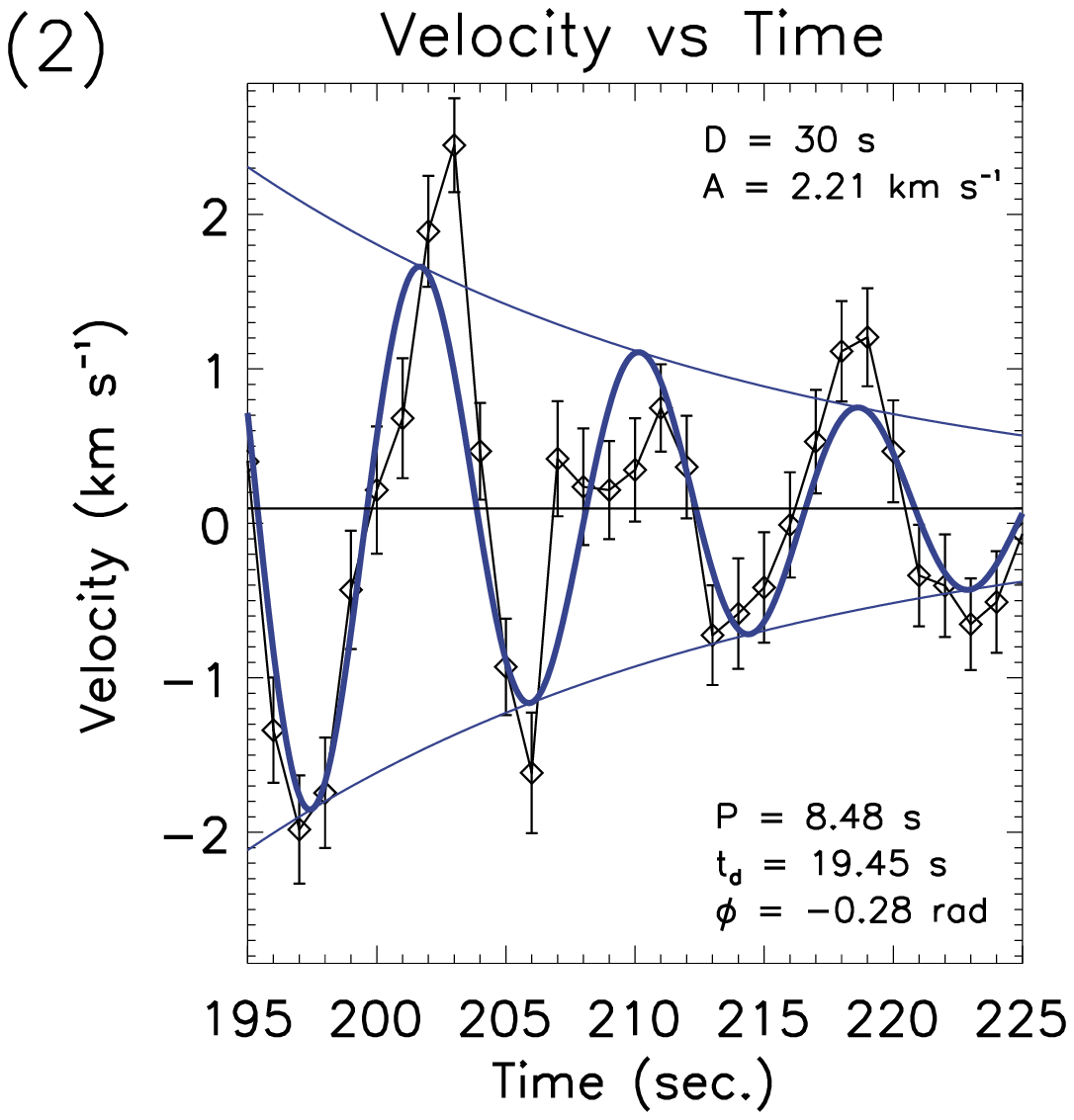}
\includegraphics[angle=90,width=6.1cm]{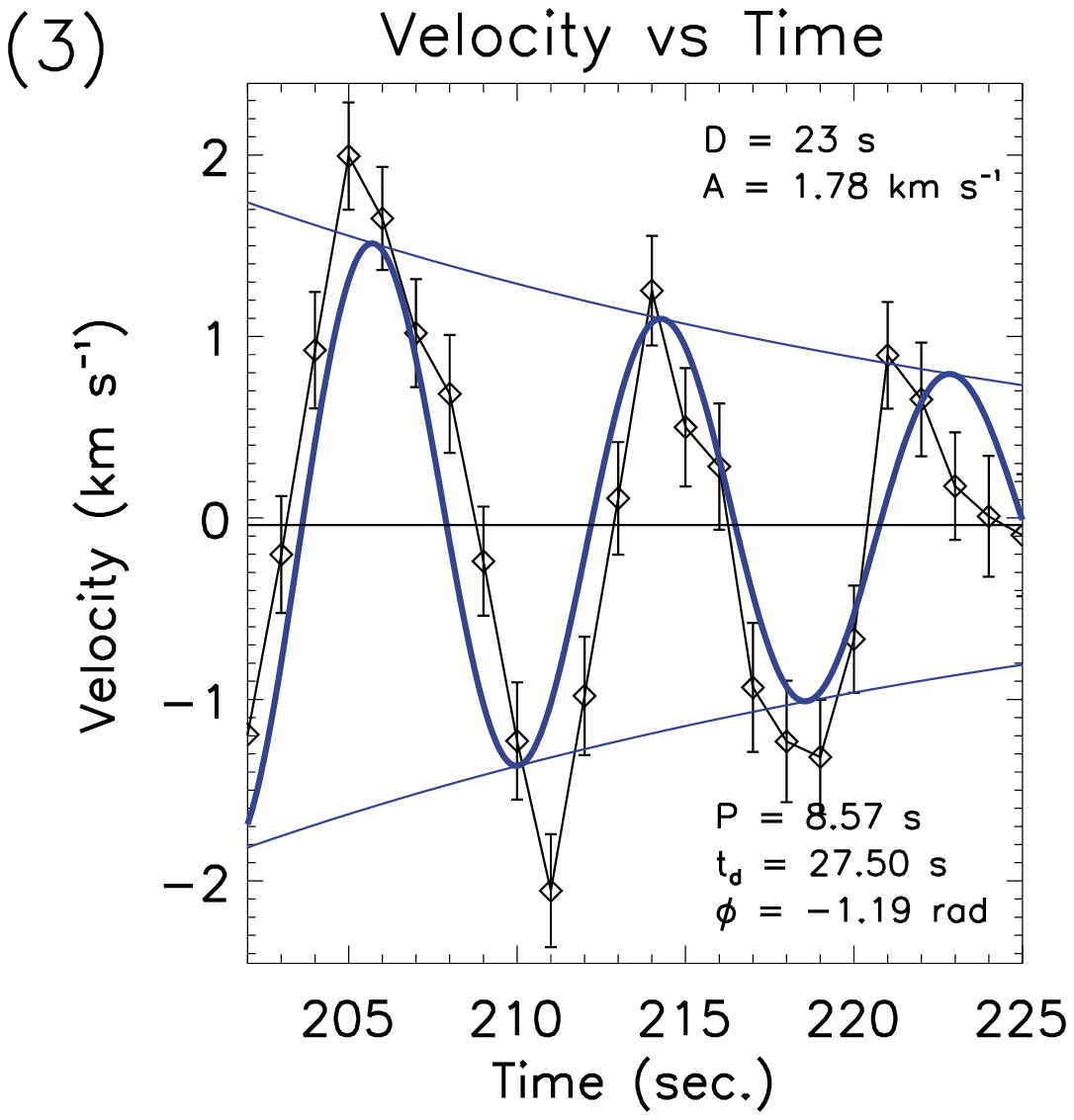}\includegraphics[angle=90,width=6.1cm]{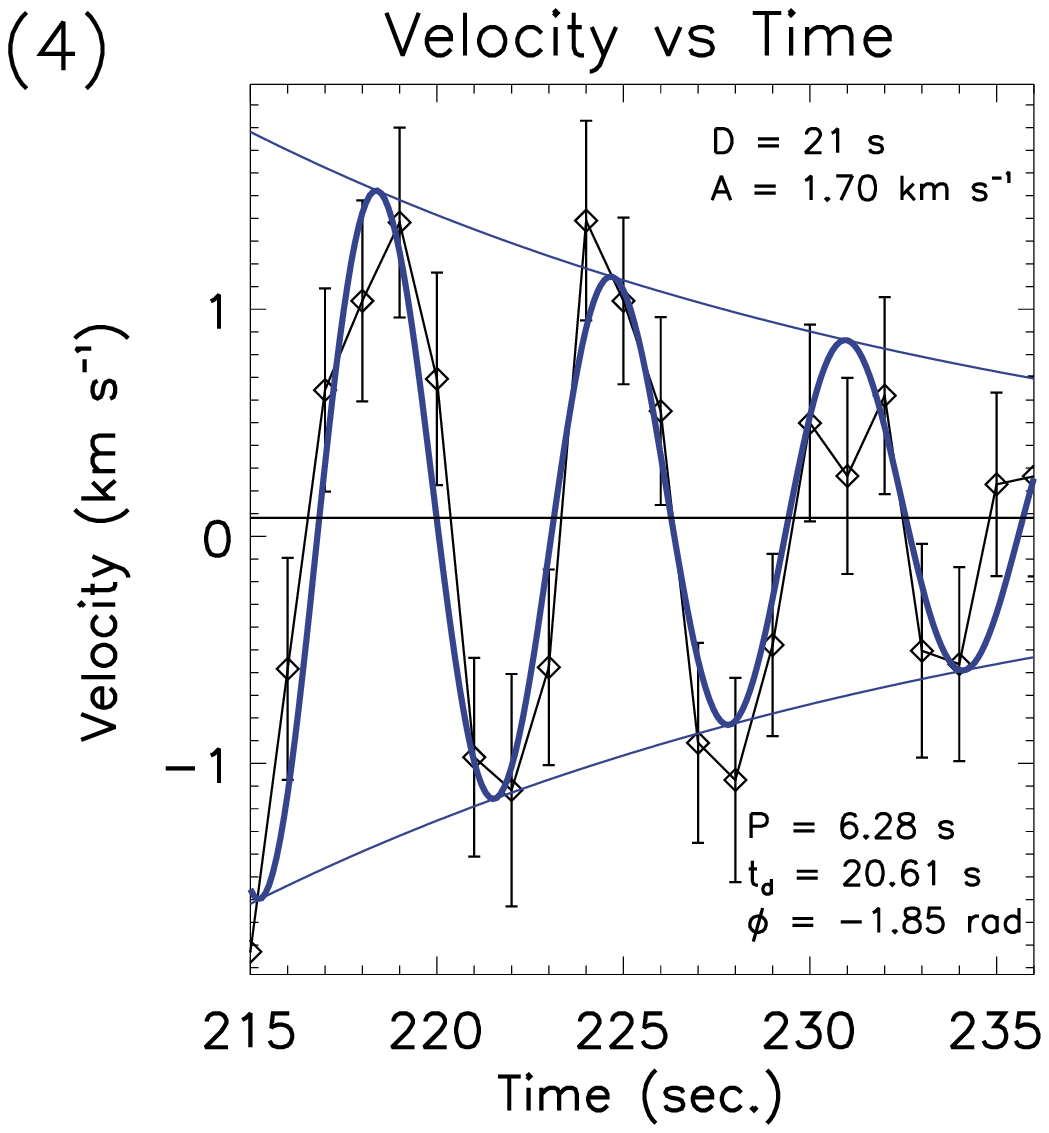}
\caption{Four events show that Doppler velocity oscillations are damping significantly with time. The location of the events (1-4) are shown by the arrow V1, V2, V3 and V4 in Figure~\ref{fig:int_xt_d}.  
The panels are similar to those in Figure~\ref{fig:int_damp}.}
\label{fig:dop_damp}
\end{figure*}
\begin{figure}
\centering
\includegraphics[angle=90,width=10.5cm]{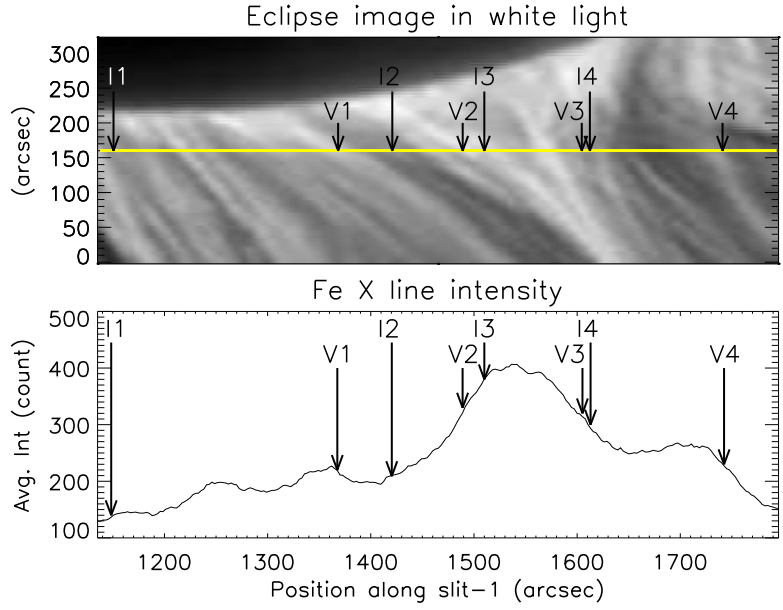}
\caption{The top and bottom panels are enlarged parts of the Figure~\ref{fig:slit_posn} and \ref{fig:int_xt}, respectively, 
used to indicate the locations of the damping events shown in Figure~\ref{fig:int_damp} and \ref{fig:dop_damp}.
Arrows I1 to I4 show the locations of four intensity damping events (1--4) respectively as shown in Figure~\ref{fig:int_damp}.
Arrows V1 to V4 show the locations of four Doppler velocity damping events (1--4), respectively, presented in Figure~\ref{fig:dop_damp}.
The damping parameters are listed in Table~\ref{tab10}.}
\label{fig:int_xt_d}
\end{figure}

One of the possible coronal heating source is the damping of MHD waves. 
To search for damping signature in the data, each time series at each location is inspected visually. 
Damped temporal samples are fitted with damped sine function (Equation~\ref{eqn1}) using the MPFIT programme in IDL. Where $A_{0}$ is the mean, A is the amplitude at time zero, 
D is the total duration, P is the period of oscillation, $\phi$ is the phase at time zero and $t_{d}$ is the damping time.
The average trend has been subtracted before fitting. 
Figure~\ref{fig:int_damp} and \ref{fig:dop_damp} show that the intensity oscillations  and 
Doppler velocity oscillations damps significantly at few locations. 
The locations of the damping are shown in Figure~\ref{fig:int_xt_d}.  
We did not find any damping signature of FWHM variations.
\begin{equation} \label{eqn1}
f(x)= A_{0}+ASin(\frac{2\pi}{P}+\phi){e}^{-\frac{t}{t_{d}}}
\end{equation}

\begin{table}
\caption{Damping Properties. Locations of the events are shown in Figure~\ref{fig:int_xt_d}.}
\begin{tabular}{ccccc}
\hline
\hline
\multicolumn{1}{c}{Parameter}&\multicolumn{1}{c}{Loaction}&\multicolumn{1}{c}{Periods}&\multicolumn{1}{c}{Damping time}&\multicolumn{1}{c}{Quality factor}\\
\hline
\hline
Intensity & I1 & 7 & 15 & 2.14 \\
 & I2 & 6 & 15 & 2.5 \\
 & I3 & 11 & 24 & 2.18 \\
 & I4 & 10 & 32 & 3.2 \\
\hline
Doppler & V1 & 9 & 15 & 1.66 \\
Velocity & V2 & 9 & 19 & 2.11 \\
 & V3 & 9 & 28 & 3.11 \\
 & V4 & 6 & 21 & 3.5 \\
\hline
\hline
\end{tabular}
\label{tab10}
\end{table}

It is likely that that damping signature of oscillations is due to MHD wave damping within coronal structures and 
they can play an important role in coronal heating.

\begin{figure*}
\centering
\includegraphics[angle=90,trim = 0mm 28mm 51mm 12mm, clip,width=9.0cm]{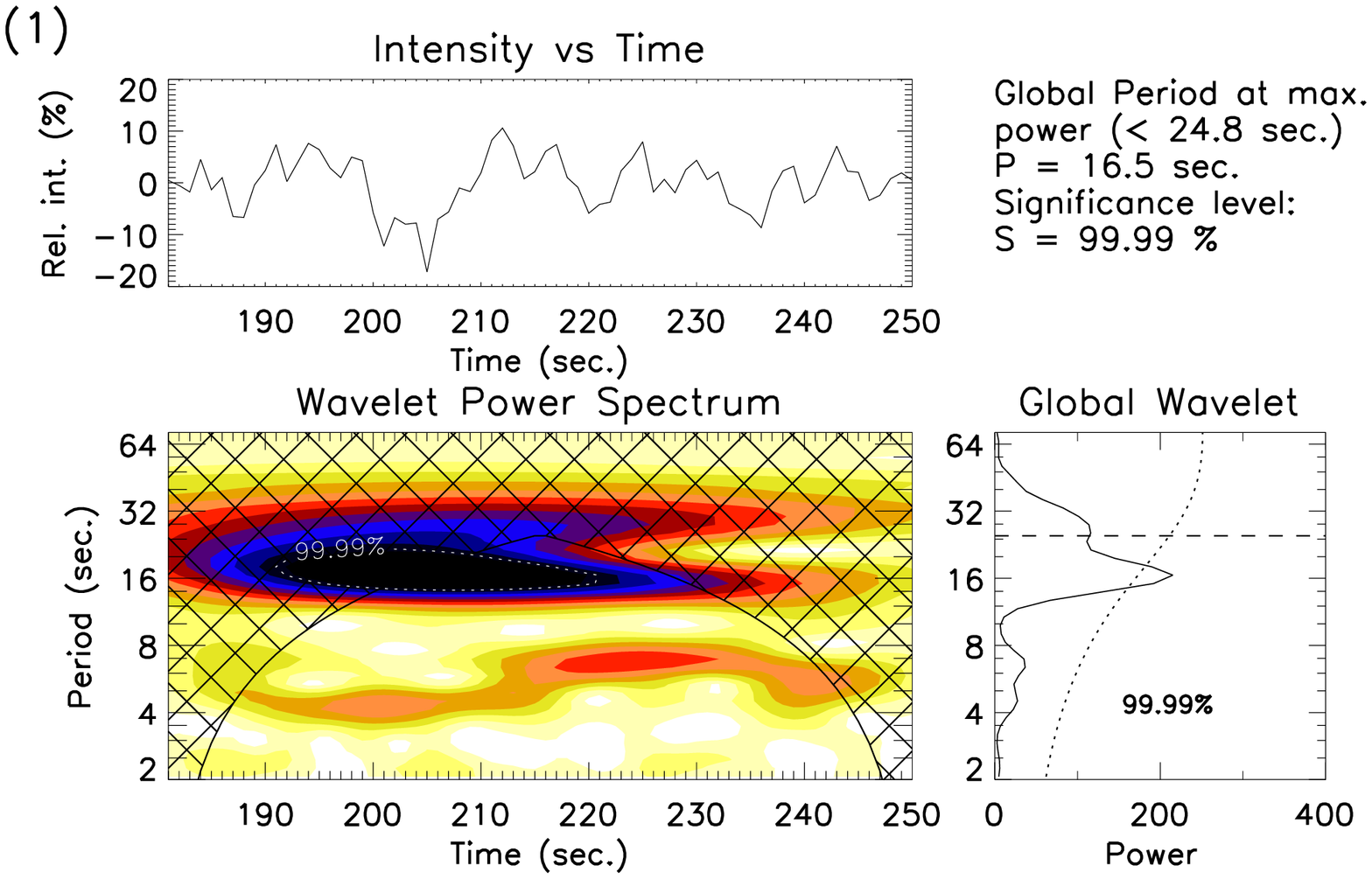}
\includegraphics[angle=90,trim = 0mm 28mm 51mm 12mm, clip,width=9.0cm]{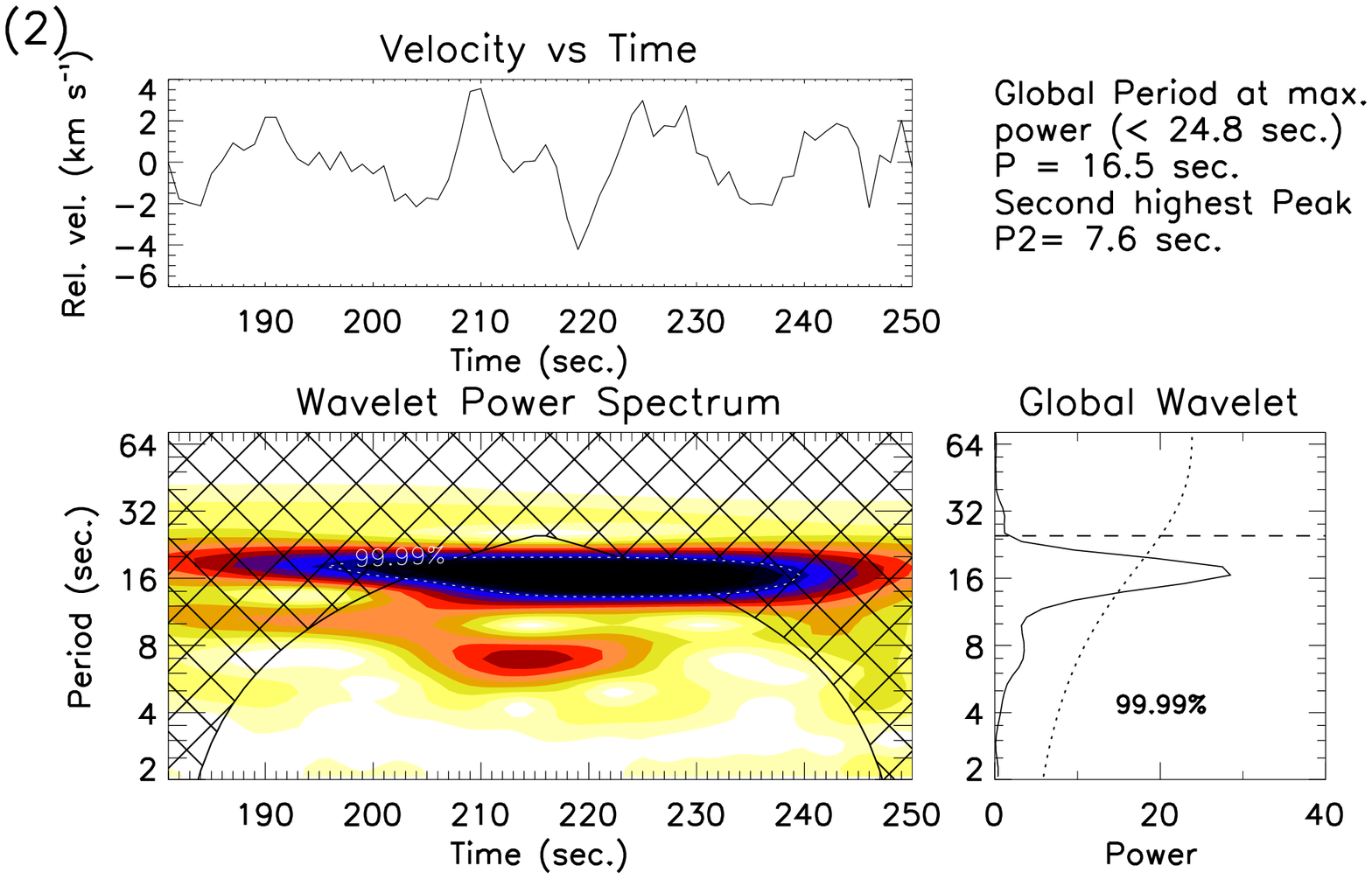}
\includegraphics[angle=90,trim = 0mm 28mm 51mm 12mm, clip,width=9.0cm]{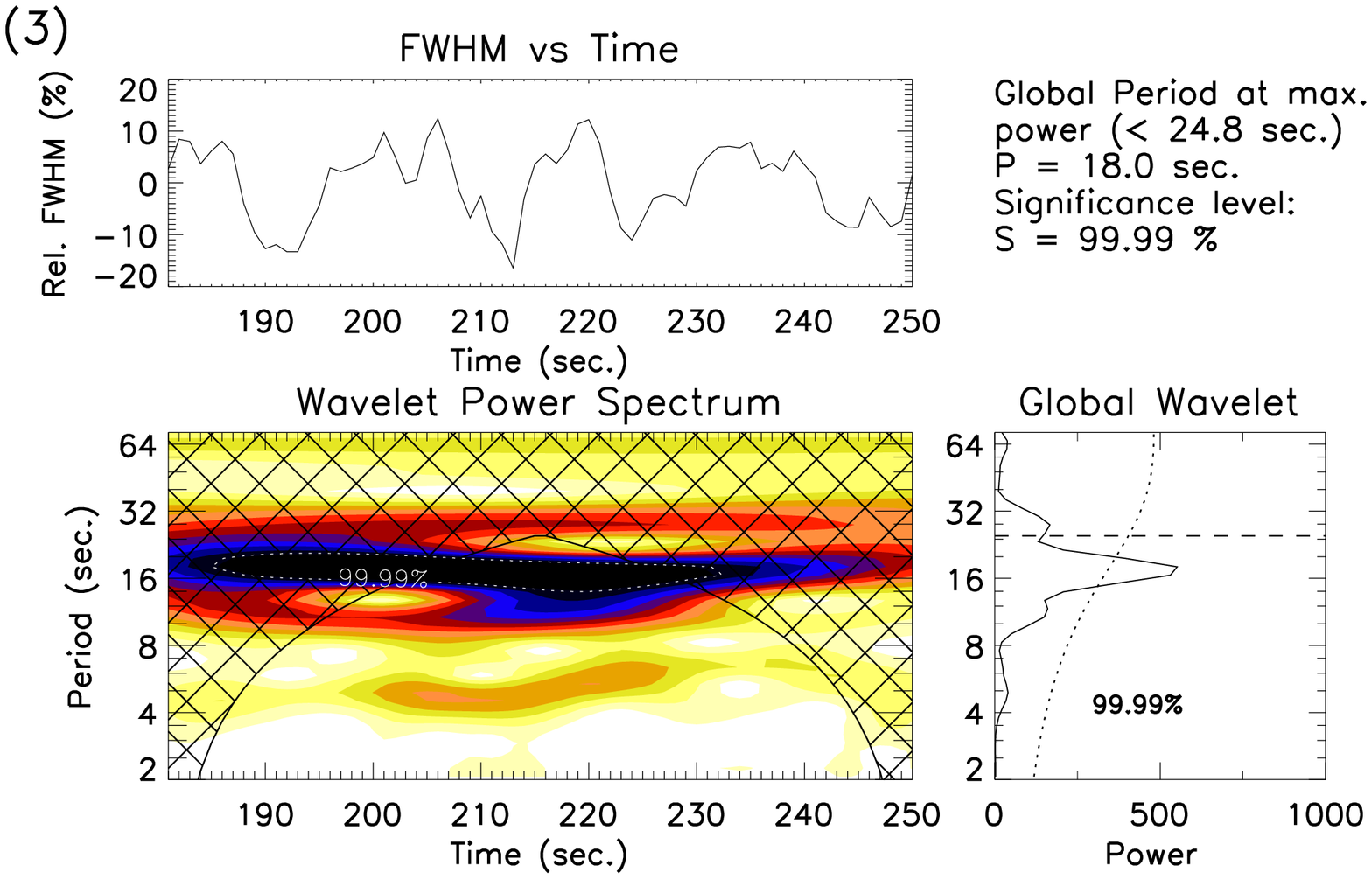}
\caption{(1) shows the result from wavelet analysis of intensity variations similar to Figure~\ref{fig:int_wavelet}. Similarly, (2) form Doppler velocity variations and (3) is from FWHM variations.   
All the light curves of different line parameters is shown from a particular location which is shown by the arrow R3 in Figure~\ref{fig:int_wid_dop}~(2).
The analysis shows that oscillations with identical periods is present in all the line parameters.} 
\label{fig:flow}
\end{figure*}

The damping parameters and the quality factor, the ratio of damping time over period, is given in Table~\ref{tab10}. 
Damping parameters can provide valuable information about physical parameters such as the electron density and filling factors etc \citep{2006RSPTA.364..417A}.
Figure \ref{fig:int_xt_d} shows that the damping of the intensity and Doppler velocity oscillations are generally occurring at locations where intensity gradients are high.
It may be resonant damping which is induced by the higher density gradient \citep{2003ApJ...598.1375A}.

\subsection{Oscillation in different line parameters with identical periods}

Coronal structures can support different MHD wave modes \citep{2005LRSP....2....3N}.
The observational signatures are different for each wave mode \citep{2010ApJ...721..744K}. 
In order to identify MHD wave modes, we have compared the time periods of the significant oscillations of the parameters at each pixel location inside the ROI.
In this case, we have assumed that while determining the periods of
the significant oscillations there can be a scatter error and we have taken it as
the nearest period (resolution) in the wavelet period spectrum.
The periods that correspond to detections in individual parameters, and their combinations, are plotted against period and slit coordinate in Figure~\ref{fig:int_wid_dop}.
According to the MHD wave theory \citep{2005LRSP....2....3N}, generally, intensity and FWHM oscillations are associated with fast sausage mode, intensity and Doppler velocity
oscillations are associated with slow acoustic mode, Doppler velocity shift and FWHM oscillations are associated with torsional Alfv\'en mode and only 
Doppler velocity shift oscillations are associated with kink mode.

In the rightmost panel of Figure~\ref{fig:int_wid_dop}~(2), we show the locations where we find oscillations in all the three line parameters with identical periods. 
A representative example is shown in Figure~\ref{fig:flow} where it shows presence of similar periodicities in different parameters. 
The correlation analysis shows that the intensity and velocity oscillations are in phase (correlation coefficient = 0.42) 
and intensity and FWHM oscillations are in opposite phase (C.C.= -0.52)
and velocity and FWHM are in opposite phase (C.C.= -0.64) as well, which might indicate a common origin of these oscillations.
The correlated oscillations between all the parameters seem be located close to boundaries of the streamer structure. 

\section{Discussion}
From an eclipse expedition and using a multi-slit spectrograph, we have studied the oscillation properties of the coronal plasma. 
We found that the intensity, Doppler velocity and width show significant oscillations with periods ranging from 6 - 25 s at many locations in the red line. 
The green line parameters also show periodic oscillations between 10 - 25 s at different locations.
These oscillations can be 
interpreted in terms of the presence of magnetohydrodynamic waves in the corona. 
Our statistical analysis shows that the intensity and Doppler velocity oscillations are more 
frequent around 15~s and 6~s but less frequent around the 10 s periods. May be the periods around 10~s are damped more effectively, making it difficult to be observed.

The power and period distribution of intensity, Doppler velocity and width 
variations (Figure~\ref{fig:int_xt},~\ref{fig:vel_xt} and \ref{fig:wid_xt}) reveal that they have a 
slight tendency to occur preferentially close to the boundaries of the structures where intensity 
gradients are relatively high. This result confirms  earlier observations \citep{2001A&A...368.1095O,2009SoPh..260..125S,2009A&A...494..355O,2010SoPh..267..305R}, 
although they were limited to intensity only.  
 \citet{2009A&A...494..355O} reported a variation of wave
mode with frequency and location. The wave mode was found to change from slow magnetoacoustic in
the plage regions to fast magnetoacoustic waves at structure
boundaries. Their analysis also shows that the ``higher-frequency
oscillations'' greater than 8 mHz occur preferentially at the edges of moss areas
which they  interpreted as  due to resonant absorption.
We should point out that if there are multiple finer structures along the line of sight, as well as across it, 
wave detections may be  difficult because signals are out of phase and polarised in different directions and 
the resultant oscillation signature may cancel out. 
Whereas, at the edges of a large structure, where the line of sight passes through fewer substructures, 
the wave signature is retained and will be easier to detect.
While performing a numerical experiment, \citet{2012ApJ...746...31D}, have demonstrated this effect. 
Thus absence of the wave signature within structures can also be due to a line of sight effect.

High-resolution spectroscopy and sufficiently high cadence have enabled us to find evidence of strong damping. 
To our knowledge, we detect the damping of the ``high-frequency oscillations'' with periods around 10 s for the first time. 
These results therefore provide additional evidence in favor of ``high-frequency waves'' damping in the corona, 
which is demonstrated by \citet{1994ApJ...435..482P,1994ApJ...435..502P} and \citet{1995SoPh..157..103L,1995SoPh..161..269L} 
to be a necessary condition for waves to heat the solar corona. 
We find that only the intensity and 
Doppler velocity oscillations are damping significantly at a few locations.
The observed intensity oscillations can be due to either the fast magnetoacoustic mode \citep{2003A&A...409..325C} 
or  the fast kink mode \citep{2008ApJ...676L..73V} if the oscillating plasma moves in and out of observing pixel position. 
The Doppler velocity oscillations are more likely  due to the fast kink mode. The fast magnetoacoustic mode can be damped by shocks \citep{1995SoPh..159..399N}.
Our analysis also shows that the damping events are generally located where intensity gradients are high.
It is possible that the kink waves are damped by resonant absorption due to the higher density gradient \citep{2002ApJ...577..475R,2002A&A...394L..39G,2003ApJ...598.1375A}.
It should be noted that the kink mode generally oscillate in transverse direction \citep{1999ApJ...520..880A}, though clear signature of oscillation in vertical
direction was also found \citep{2004A&A...421L..33W}.
Hence, they have a different plane of oscillation (different polarization) and can cancel out the signal at the bright overlapping loops.
This effect also can contribute to the detection of damping close to the boundaries where line-of-sight signal passes through fewer substructures but inside the bright structure it passes through multiple loops.

The periods and damping times of MHD modes are dependent on the plasma densities inside and surroundings of the oscillating structures.
Damping parameters provide information about physical parameters such as the electron density, filling factors {\it etc.} \citep{2000SoPh..193..139R,2006RSPTA.364..417A}.
As the fast kink mode has relatively shorter damping times and hence it is rarely detectable. 
The ratio between periods and damping times, the quality factor, of these waves can help to understand the damping mechanism \citep{2002ApJ...576L.153O,2003ApJ...598.1375A}.
It also gives an idea about the relative strength of the damping.
Though, quality factor greater than 0.5 signifies damping,  
statistical observational studies have reported that it is varying from 0.6 to 5.4 (\citet{2013A&A...552A.138V} and references therein).
In our observation, the quality factor varies from 1.6 to 3.5, which
means that our observed damping oscillations lie in the under-damped regime. Hence, it could be an energy source for heating. 
Due to availability of limited data we could not detect damping at several locations. 
It is to be noted that to detect damping of shorter periodicities one needs very high cadence observations.

We also try to characterize the nature of the wave modes.
The oscillations in intensity and velocity can be interpreted as due to compressional waves, 
whereas oscillations with a shared period in intensity and FWHM can be due to the propagation of the fast sausage mode.
The Doppler velocity and FWHM oscillations can be associated with torsional Alfv\'en mode whereas the existence Doppler velocity oscillation only can be attributed 
to fast kink mode (see \citet{2005LRSP....2....3N,2006RSPTA.364..417A}).

One may also interpret the correlated oscillations between intensity, Doppler velocity, and line width as shown in Figure \ref{fig:flow} 
as due to presence of quasi-periodic flows \citep{2010ApJ...722.1013D,2011ApJ...727L..37T,2012ApJ...759..144T}. 
Close to the streamer boundaries we do see such coherent oscillations and the scenario of the presence of flows can not be ruled out. 
Although, numerical simulation of \citet{2010ApJ...724L.194V} demonstrated that due to the in-phase behaviour of velocity and density 
perturbations, upward propagating waves can cause similar effects.

If the selected locations correspond to several wave guides supporting different wave modes it may not be possible to isolate them and identify them. 
Thus it is not surprising that we see different signatures at different locations. 
Statistically which are the significant modes and how they play a role in coronal heating are key questions.

\section*{Acknowledgments}
We would like to thank Professor S.S. Hasan for encouragement and support and Dr. Krishna Prasad S. for valuable insight during the data analysis process. 
We thank the anonymous referee for his/her valuable comments which has enabled us to improve the presentation and the manuscript.
We thank F. Gabriel and K. Ravi for their help during the fabrication, alignment, installation and testing of the setup at the eclipse site. We also like to thank
V. Natarajan for his help in making the schematic diagram of the optical layout.

\bibliographystyle{spr-mp-sola.bst}
\bibliography{references}

\begin{thebibliography}{47}
\ifx\bisbn     \undefined \def\bisbn  #1{ISBN #1}\fi
\ifx\binits    \undefined \def\binits#1{#1}\fi
\ifx\bauthor   \undefined \def\bauthor#1{#1}\fi
\ifx\batitle   \undefined \def\batitle#1{#1}\fi
\ifx\bjtitle   \undefined \def\bjtitle#1{\textit{#1}}\fi
\ifx\bvolume   \undefined \def\bvolume#1{\textbf{#1}}\fi
\ifx\byear     \undefined \def\byear#1{#1}\fi
\ifx\bissue    \undefined \def\bissue#1{#1}\fi
\ifx\bfpage    \undefined \def\bfpage#1{#1}\fi
\ifx\blpage    \undefined \def\blpage #1{#1}\fi
\ifx\burl      \undefined \def\burl#1{\textsf{#1}}\fi
\ifx\href      \undefined \def\href#1#2{\textsf{#2}}\fi
\ifx\betal     \undefined \def\betal{\textit{et al.}}\fi
\ifx\bctitle   \undefined \def\bctitle#1{#1}\fi
\ifx\beditor   \undefined \def\beditor#1{#1}\fi
\ifx\bbtitle   \undefined \def\bbtitle#1{\textit{#1}}\fi
\ifx\bedition  \undefined \def\bedition#1{#1}\fi
\ifx\bseriesno \undefined \def\bseriesno#1{\textbf{#1}}\fi
\ifx\blocation \undefined \def\blocation#1{#1}\fi
\ifx\bsertitle \undefined \def\bsertitle#1{\textit{#1}}\fi
\ifx\bsnm      \undefined \def\bsnm#1{#1}\fi
\ifx\bsuffix   \undefined \def\bsuffix#1{#1}\fi
\ifx\bparticle \undefined \def\bparticle#1{#1}\fi
\ifx\barticle  \undefined \def\barticle#1{}\fi
\ifx\binstitute  \undefined \def\binstitute#1{#1}\fi
\ifx\bpublisher  \undefined \def\bpublisher#1{#1}\fi
\ifx\doiurl    \undefined
  \def\doiurl#1{\href{http://dx.doi.org/#1}{\textsf{DOI}}}\fi
\ifx\arxivurl  \undefined
  \def\arxivurl#1{\href{http://arxiv.org/abs/#1}{\textsf{arXiv}}}\fi
\ifx\adsurl    \undefined
  \def\adsurl#1{\href{http://adsabs.harvard.edu/abs/#1}{\textsf{ADS}}}\fi
\ifx\botherref \undefined \def\botherref#1{}\fi
\ifx\url       \undefined \def\url#1{\textsf{#1}}\fi
\ifx\bchapter  \undefined \def\bchapter#1{}\fi
\ifx\bbook     \undefined \def\bbook#1{}\fi
\ifx\bcomment  \undefined \def\bcomment#1{#1}\fi
\ifx\oauthor   \undefined \def\oauthor#1{#1}\fi
\ifx\citeauthoryear \undefined\def \citeauthoryear#1{#1}\fi
\def\endbibitem {}
\ifx\bconflocation  \undefined \def\bconflocation#1{#1} \fi

\bibitem[\protect\citeauthoryear{{Aschwanden}}{1987}]{1987SoPh..111..113A}
\begin{barticle}
\bauthor{\bsnm{{Aschwanden}}, \binits{M.J.}}:
\byear{1987},
\batitle{{Theory of radio pulsations in coronal loops}}.
\bjtitle{\solphys}
\bvolume{111},
\bfpage{113}.
\doiurl{10.1007/BF00145445}.
\adsurl{1987SoPh..111..113A}.
\end{barticle}
\endbibitem

\bibitem[\protect\citeauthoryear{{Aschwanden}}{2006}]{2006RSPTA.364..417A}
\begin{barticle}
\bauthor{\bsnm{{Aschwanden}}, \binits{M.J.}}:
\byear{2006},
\batitle{{Coronal magnetohydrodynamic waves and oscillations: observations and
  quests}}.
\bjtitle{Roy. Soc. London Phil. Trans. Ser. A}
\bvolume{364},
\bfpage{417}.
\doiurl{10.1098/rsta.2005.1707}.
\adsurl{2006RSPTA.364..417A}.
\end{barticle}
\endbibitem

\bibitem[\protect\citeauthoryear{{Aschwanden}
  \textit{et~al.}}{1999}]{1999ApJ...520..880A}
\begin{barticle}
\bauthor{\bsnm{{Aschwanden}}, \binits{M.J.}},
\bauthor{\bsnm{{Fletcher}}, \binits{L.}},
\bauthor{\bsnm{{Schrijver}}, \binits{C.J.}},
\bauthor{\bsnm{{Alexander}}, \binits{D.}}:
\byear{1999},
\batitle{{Coronal Loop Oscillations Observed with the Transition Region and
  Coronal Explorer}}.
\bjtitle{\apj}
\bvolume{520},
\bfpage{880}.
\doiurl{10.1086/307502}.
\adsurl{1999ApJ...520..880A}.
\end{barticle}
\endbibitem

\bibitem[\protect\citeauthoryear{{Aschwanden}
  \textit{et~al.}}{2003}]{2003ApJ...598.1375A}
\begin{barticle}
\bauthor{\bsnm{{Aschwanden}}, \binits{M.J.}},
\bauthor{\bsnm{{Nightingale}}, \binits{R.W.}},
\bauthor{\bsnm{{Andries}}, \binits{J.}},
\bauthor{\bsnm{{Goossens}}, \binits{M.}},
\bauthor{\bsnm{{Van Doorsselaere}}, \binits{T.}}:
\byear{2003},
\batitle{{Observational Tests of Damping by Resonant Absorption in Coronal Loop
  Oscillations}}.
\bjtitle{\apj}
\bvolume{598},
\bfpage{1375}.
\doiurl{10.1086/379104}.
\adsurl{2003ApJ...598.1375A}.
\end{barticle}
\endbibitem

\bibitem[\protect\citeauthoryear{{Banerjee}
  \textit{et~al.}}{2007}]{2007SoPh..246....3B}
\begin{barticle}
\bauthor{\bsnm{{Banerjee}}, \binits{D.}},
\bauthor{\bsnm{{Erd{\'e}lyi}}, \binits{R.}},
\bauthor{\bsnm{{Oliver}}, \binits{R.}},
\bauthor{\bsnm{{O'Shea}}, \binits{E.}}:
\byear{2007},
\batitle{{Present and Future Observing Trends in Atmospheric
  Magnetoseismology}}.
\bjtitle{\solphys}
\bvolume{246},
\bfpage{3}.
\doiurl{10.1007/s11207-007-9029-z}.
\adsurl{2007SoPh..246....3B}.
\end{barticle}
\endbibitem

\bibitem[\protect\citeauthoryear{{Billings}}{1959}]{1959ApJ...130..215B}
\begin{barticle}
\bauthor{\bsnm{{Billings}}, \binits{D.E.}}:
\byear{1959},
\batitle{{Velocity Fields in a Coronal Region with a Possible Hydromagnetic
  Interpretation.}}
\bjtitle{\apj}
\bvolume{130},
\bfpage{215}.
\doiurl{10.1086/146710}.
\adsurl{1959ApJ...130..215B}.
\end{barticle}
\endbibitem

\bibitem[\protect\citeauthoryear{{Cooper}, {Nakariakov}, and
  {Williams}}{2003}]{2003A&A...409..325C}
\begin{barticle}
\bauthor{\bsnm{{Cooper}}, \binits{F.C.}},
\bauthor{\bsnm{{Nakariakov}}, \binits{V.M.}},
\bauthor{\bsnm{{Williams}}, \binits{D.R.}}:
\byear{2003},
\batitle{{Short period fast waves in solar coronal loops}}.
\bjtitle{\aap}
\bvolume{409},
\bfpage{325}.
\doiurl{10.1051/0004-6361:20031071}.
\adsurl{2003A\%26A...409..325C}.
\end{barticle}
\endbibitem

\bibitem[\protect\citeauthoryear{{Cowsik}
  \textit{et~al.}}{1999}]{1999SoPh..188...89C}
\begin{barticle}
\bauthor{\bsnm{{Cowsik}}, \binits{R.}},
\bauthor{\bsnm{{Singh}}, \binits{J.}},
\bauthor{\bsnm{{Saxena}}, \binits{A.K.}},
\bauthor{\bsnm{{Srinivasan}}, \binits{R.}},
\bauthor{\bsnm{{Raveendran}}, \binits{A.V.}}:
\byear{1999},
\batitle{{Short-period intensity oscillations in the solar corona observed
  during the total solar eclipse of 26 February 1998}}.
\bjtitle{\solphys}
\bvolume{188},
\bfpage{89}.
\adsurl{1999SoPh..188...89C}.
\end{barticle}
\endbibitem

\bibitem[\protect\citeauthoryear{{De Moortel} and
  {Pascoe}}{2012}]{2012ApJ...746...31D}
\begin{barticle}
\bauthor{\bsnm{{De Moortel}}, \binits{I.}},
\bauthor{\bsnm{{Pascoe}}, \binits{D.J.}}:
\byear{2012},
\batitle{{The Effects of Line-of-sight Integration on Multistrand Coronal Loop
  Oscillations}}.
\bjtitle{\apj}
\bvolume{746},
\bfpage{31}.
\doiurl{10.1088/0004-637X/746/1/31}.
\adsurl{2012ApJ...746...31D}.
\end{barticle}
\endbibitem

\bibitem[\protect\citeauthoryear{{De Pontieu} and
  {McIntosh}}{2010}]{2010ApJ...722.1013D}
\begin{barticle}
\bauthor{\bsnm{{De Pontieu}}, \binits{B.}},
\bauthor{\bsnm{{McIntosh}}, \binits{S.W.}}:
\byear{2010},
\batitle{{Quasi-periodic Propagating Signals in the Solar Corona: The Signature
  of Magnetoacoustic Waves or High-velocity Upflows?}}
\bjtitle{\apj}
\bvolume{722},
\bfpage{1013}.
\doiurl{10.1088/0004-637X/722/2/1013}.
\adsurl{2010ApJ...722.1013D}.
\end{barticle}
\endbibitem

\bibitem[\protect\citeauthoryear{{Goossens}, {Andries}, and
  {Aschwanden}}{2002}]{2002A&A...394L..39G}
\begin{barticle}
\bauthor{\bsnm{{Goossens}}, \binits{M.}},
\bauthor{\bsnm{{Andries}}, \binits{J.}},
\bauthor{\bsnm{{Aschwanden}}, \binits{M.J.}}:
\byear{2002},
\batitle{{Coronal loop oscillations. An interpretation in terms of resonant
  absorption of quasi-mode kink oscillations}}.
\bjtitle{\aap}
\bvolume{394},
\bfpage{L39}.
\doiurl{10.1051/0004-6361:20021378}.
\adsurl{2002A\%26A...394L..39G}.
\end{barticle}
\endbibitem

\bibitem[\protect\citeauthoryear{{Habbal}
  \textit{et~al.}}{2011}]{2011ApJ...734..120H}
\begin{barticle}
\bauthor{\bsnm{{Habbal}}, \binits{S.R.}},
\bauthor{\bsnm{{Druckm{\"u}ller}}, \binits{M.}},
\bauthor{\bsnm{{Morgan}}, \binits{H.}},
\bauthor{\bsnm{{Ding}}, \binits{A.}},
\bauthor{\bsnm{{Johnson}}, \binits{J.}},
\bauthor{\bsnm{{Druckm{\"u}llerov{\'a}}}, \binits{H.}},
\bauthor{\bsnm{{Daw}}, \binits{A.}},
\bauthor{\bsnm{{Arndt}}, \binits{M.B.}},
\bauthor{\bsnm{{Dietzel}}, \binits{M.}},
\bauthor{\bsnm{{Saken}}, \binits{J.}}:
\byear{2011},
\batitle{{Thermodynamics of the Solar Corona and Evolution of the Solar
  Magnetic Field as Inferred from the Total Solar Eclipse Observations of 2010
  July 11}}.
\bjtitle{\apj}
\bvolume{734},
\bfpage{120}.
\doiurl{10.1088/0004-637X/734/2/120}.
\adsurl{2011ApJ...734..120H}.
\end{barticle}
\endbibitem

\bibitem[\protect\citeauthoryear{{Katsiyannis}
  \textit{et~al.}}{2003}]{2003A&A...406..709K}
\begin{barticle}
\bauthor{\bsnm{{Katsiyannis}}, \binits{A.C.}},
\bauthor{\bsnm{{Williams}}, \binits{D.R.}},
\bauthor{\bsnm{{McAteer}}, \binits{R.T.J.}},
\bauthor{\bsnm{{Gallagher}}, \binits{P.T.}},
\bauthor{\bsnm{{Keenan}}, \binits{F.P.}},
\bauthor{\bsnm{{Murtagh}}, \binits{F.}}:
\byear{2003},
\batitle{{Eclipse observations of high-frequency oscillations in active region
  coronal loops}}.
\bjtitle{\aap}
\bvolume{406},
\bfpage{709}.
\doiurl{10.1051/0004-6361:20030458}.
\adsurl{2003A\%26A...406..709K}.
\end{barticle}
\endbibitem

\bibitem[\protect\citeauthoryear{{Kitagawa}
  \textit{et~al.}}{2010}]{2010ApJ...721..744K}
\begin{barticle}
\bauthor{\bsnm{{Kitagawa}}, \binits{N.}},
\bauthor{\bsnm{{Yokoyama}}, \binits{T.}},
\bauthor{\bsnm{{Imada}}, \binits{S.}},
\bauthor{\bsnm{{Hara}}, \binits{H.}}:
\byear{2010},
\batitle{{Mode Identification of MHD Waves in an Active Region Observed with
  Hinode/EIS}}.
\bjtitle{\apj}
\bvolume{721},
\bfpage{744}.
\doiurl{10.1088/0004-637X/721/1/744}.
\adsurl{2010ApJ...721..744K}.
\end{barticle}
\endbibitem

\bibitem[\protect\citeauthoryear{{Koutchmy}, {Zhugzhda}, and
  {Locans}}{1983}]{1983A&A...120..185K}
\begin{barticle}
\bauthor{\bsnm{{Koutchmy}}, \binits{S.}},
\bauthor{\bsnm{{Zhugzhda}}, \binits{I.D.}},
\bauthor{\bsnm{{Locans}}, \binits{V.}}:
\byear{1983},
\batitle{{Short period coronal oscillations - Observation and interpretation}}.
\bjtitle{\aap}
\bvolume{120},
\bfpage{185}.
\adsurl{1983A\%26A...120..185K}.
\end{barticle}
\endbibitem

\bibitem[\protect\citeauthoryear{{Laing} and
  {Edwin}}{1995a}]{1995SoPh..161..269L}
\begin{barticle}
\bauthor{\bsnm{{Laing}}, \binits{G.B.}},
\bauthor{\bsnm{{Edwin}}, \binits{P.M.}}:
\byear{1995}a,
\batitle{{Dissipating the Energy of Magnetoacoustic Waves in a Structured
  Atmosphere}}.
\bjtitle{\solphys}
\bvolume{161},
\bfpage{269}.
\doiurl{10.1007/BF00732071}.
\adsurl{1995SoPh..161..269L}.
\end{barticle}
\endbibitem

\bibitem[\protect\citeauthoryear{{Laing} and
  {Edwin}}{1995b}]{1995SoPh..157..103L}
\begin{barticle}
\bauthor{\bsnm{{Laing}}, \binits{G.B.}},
\bauthor{\bsnm{{Edwin}}, \binits{P.M.}}:
\byear{1995}b,
\batitle{{Dissipating waves in a hot, compressible, magnetic, structured
  atmosphere}}.
\bjtitle{\solphys}
\bvolume{157},
\bfpage{103}.
\doiurl{10.1007/BF00680611}.
\adsurl{1995SoPh..157..103L}.
\end{barticle}
\endbibitem

\bibitem[\protect\citeauthoryear{{Morton} and
  {McLaughlin}}{2013}]{2013A&A...553L..10M}
\begin{barticle}
\bauthor{\bsnm{{Morton}}, \binits{R.J.}},
\bauthor{\bsnm{{McLaughlin}}, \binits{J.A.}}:
\byear{2013},
\batitle{{Hi-C and AIA observations of transverse magnetohydrodynamic waves in
  active regions}}.
\bjtitle{\aap}
\bvolume{553},
\bfpage{L10}.
\doiurl{10.1051/0004-6361/201321465}.
\adsurl{2013A\%26A...553L..10M}.
\end{barticle}
\endbibitem

\bibitem[\protect\citeauthoryear{{Nakariakov} and
  {Roberts}}{1995}]{1995SoPh..159..399N}
\begin{barticle}
\bauthor{\bsnm{{Nakariakov}}, \binits{V.M.}},
\bauthor{\bsnm{{Roberts}}, \binits{B.}}:
\byear{1995},
\batitle{{On Fast Magnetosonic Coronal Pulsations}}.
\bjtitle{\solphys}
\bvolume{159},
\bfpage{399}.
\doiurl{10.1007/BF00686541}.
\adsurl{1995SoPh..159..399N}.
\end{barticle}
\endbibitem

\bibitem[\protect\citeauthoryear{{Nakariakov} and
  {Verwichte}}{2005}]{2005LRSP....2....3N}
\begin{barticle}
\bauthor{\bsnm{{Nakariakov}}, \binits{V.M.}},
\bauthor{\bsnm{{Verwichte}}, \binits{E.}}:
\byear{2005},
\batitle{{Coronal Waves and Oscillations}}.
\bjtitle{Living Rev. Solar Phys.}
\bvolume{2},
\bfpage{3}.
\adsurl{2005LRSP....2....3N}.
\end{barticle}
\endbibitem

\bibitem[\protect\citeauthoryear{{Ofman} and
  {Aschwanden}}{2002}]{2002ApJ...576L.153O}
\begin{barticle}
\bauthor{\bsnm{{Ofman}}, \binits{L.}},
\bauthor{\bsnm{{Aschwanden}}, \binits{M.J.}}:
\byear{2002},
\batitle{{Damping Time Scaling of Coronal Loop Oscillations Deduced from
  Transition Region and Coronal Explorer Observations}}.
\bjtitle{\apjl}
\bvolume{576},
\bfpage{L153}.
\doiurl{10.1086/343886}.
\adsurl{2002ApJ...576L.153O}.
\end{barticle}
\endbibitem

\bibitem[\protect\citeauthoryear{{O'Shea} and
  {Doyle}}{2009}]{2009A&A...494..355O}
\begin{barticle}
\bauthor{\bsnm{{O'Shea}}, \binits{E.}},
\bauthor{\bsnm{{Doyle}}, \binits{J.G.}}:
\byear{2009},
\batitle{{On oscillations found in an active region with EIS on Hinode}}.
\bjtitle{\aap}
\bvolume{494},
\bfpage{355}.
\doiurl{10.1051/0004-6361:200810916}.
\adsurl{2009A\%26A...494..355O}.
\end{barticle}
\endbibitem

\bibitem[\protect\citeauthoryear{{O'Shea}
  \textit{et~al.}}{2001}]{2001A&A...368.1095O}
\begin{barticle}
\bauthor{\bsnm{{O'Shea}}, \binits{E.}},
\bauthor{\bsnm{{Banerjee}}, \binits{D.}},
\bauthor{\bsnm{{Doyle}}, \binits{J.G.}},
\bauthor{\bsnm{{Fleck}}, \binits{B.}},
\bauthor{\bsnm{{Murtagh}}, \binits{F.}}:
\byear{2001},
\batitle{{Active region oscillations}}.
\bjtitle{\aap}
\bvolume{368},
\bfpage{1095}.
\doiurl{10.1051/0004-6361:20010073}.
\adsurl{2001A\%26A...368.1095O}.
\end{barticle}
\endbibitem

\bibitem[\protect\citeauthoryear{{Pasachoff}
  \textit{et~al.}}{2002}]{2002SoPh..207..241P}
\begin{barticle}
\bauthor{\bsnm{{Pasachoff}}, \binits{J.M.}},
\bauthor{\bsnm{{Babcock}}, \binits{B.A.}},
\bauthor{\bsnm{{Russell}}, \binits{K.D.}},
\bauthor{\bsnm{{Seaton}}, \binits{D.B.}}:
\byear{2002},
\batitle{{Short-Period Waves That Heat the Corona Detected at the 1999
  Eclipse}}.
\bjtitle{\solphys}
\bvolume{207},
\bfpage{241}.
\doiurl{10.1023/A:1016297800478}.
\adsurl{2002SoPh..207..241P}.
\end{barticle}
\endbibitem

\bibitem[\protect\citeauthoryear{{Phillips}
  \textit{et~al.}}{2000}]{2000SoPh..193..259P}
\begin{barticle}
\bauthor{\bsnm{{Phillips}}, \binits{K.J.H.}},
\bauthor{\bsnm{{Read}}, \binits{P.D.}},
\bauthor{\bsnm{{Gallagher}}, \binits{P.T.}},
\bauthor{\bsnm{{Keenan}}, \binits{F.P.}},
\bauthor{\bsnm{{Rudawy}}, \binits{P.}},
\bauthor{\bsnm{{Rompolt}}, \binits{B.}},
\bauthor{\bsnm{{Berlicki}}, \binits{A.}},
\bauthor{\bsnm{{Buczylko}}, \binits{A.}},
\bauthor{\bsnm{{Diego}}, \binits{F.}},
\bauthor{\bsnm{{Barnsley}}, \binits{R.}},
\bauthor{\bsnm{{Smartt}}, \binits{R.N.}},
\bauthor{\bsnm{{Pasachoff}}, \binits{J.M.}},
\bauthor{\bsnm{{Babcock}}, \binits{B.A.}}:
\byear{2000},
\batitle{{SECIS: The Solar Eclipse Coronal Eclipse Imaging System}}.
\bjtitle{\solphys}
\bvolume{193},
\bfpage{259}.
\doiurl{10.1023/A:1005274827585}.
\adsurl{2000SoPh..193..259P}.
\end{barticle}
\endbibitem

\bibitem[\protect\citeauthoryear{{Porter}, {Klimchuk}, and
  {Sturrock}}{1994a}]{1994ApJ...435..502P}
\begin{barticle}
\bauthor{\bsnm{{Porter}}, \binits{L.J.}},
\bauthor{\bsnm{{Klimchuk}}, \binits{J.A.}},
\bauthor{\bsnm{{Sturrock}}, \binits{P.A.}}:
\byear{1994}a,
\batitle{{The possible role of high-frequency waves in heating solar coronal
  loops}}.
\bjtitle{\apj}
\bvolume{435},
\bfpage{502}.
\doiurl{10.1086/174831}.
\adsurl{1994ApJ...435..502P}.
\end{barticle}
\endbibitem

\bibitem[\protect\citeauthoryear{{Porter}, {Klimchuk}, and
  {Sturrock}}{1994b}]{1994ApJ...435..482P}
\begin{barticle}
\bauthor{\bsnm{{Porter}}, \binits{L.J.}},
\bauthor{\bsnm{{Klimchuk}}, \binits{J.A.}},
\bauthor{\bsnm{{Sturrock}}, \binits{P.A.}}:
\byear{1994}b,
\batitle{{The possible role of MHD waves in heating the solar corona}}.
\bjtitle{\apj}
\bvolume{435},
\bfpage{482}.
\doiurl{10.1086/174830}.
\adsurl{1994ApJ...435..482P}.
\end{barticle}
\endbibitem

\bibitem[\protect\citeauthoryear{{Roberts}}{2000}]{2000SoPh..193..139R}
\begin{barticle}
\bauthor{\bsnm{{Roberts}}, \binits{B.}}:
\byear{2000},
\batitle{{Waves and Oscillations in the Corona - (Invited Review)}}.
\bjtitle{\solphys}
\bvolume{193},
\bfpage{139}.
\doiurl{10.1023/A:1005237109398}.
\adsurl{2000SoPh..193..139R}.
\end{barticle}
\endbibitem

\bibitem[\protect\citeauthoryear{{Rudawy}
  \textit{et~al.}}{2010}]{2010SoPh..267..305R}
\begin{barticle}
\bauthor{\bsnm{{Rudawy}}, \binits{P.}},
\bauthor{\bsnm{{Phillips}}, \binits{K.J.H.}},
\bauthor{\bsnm{{Buczylko}}, \binits{A.}},
\bauthor{\bsnm{{Williams}}, \binits{D.R.}},
\bauthor{\bsnm{{Keenan}}, \binits{F.P.}}:
\byear{2010},
\batitle{{Search for Rapid Changes in the Visible-Light Corona during the 21
  June 2001 Total Solar Eclipse}}.
\bjtitle{\solphys}
\bvolume{267},
\bfpage{305}.
\doiurl{10.1007/s11207-010-9647-8}.
\adsurl{2010SoPh..267..305R}.
\end{barticle}
\endbibitem

\bibitem[\protect\citeauthoryear{{Ruderman} and
  {Roberts}}{2002}]{2002ApJ...577..475R}
\begin{barticle}
\bauthor{\bsnm{{Ruderman}}, \binits{M.S.}},
\bauthor{\bsnm{{Roberts}}, \binits{B.}}:
\byear{2002},
\batitle{{The Damping of Coronal Loop Oscillations}}.
\bjtitle{\apj}
\bvolume{577},
\bfpage{475}.
\doiurl{10.1086/342130}.
\adsurl{2002ApJ...577..475R}.
\end{barticle}
\endbibitem

\bibitem[\protect\citeauthoryear{{Ru{\v s}in} and
  {Minarovjech}}{1994}]{1994scs..conf..487R}
\begin{bchapter}
\bauthor{\bsnm{{Ru{\v s}in}}, \binits{V.}},
\bauthor{\bsnm{{Minarovjech}}, \binits{M.}}:
\byear{1994},
\bctitle{{Detection of small-scale dynamics in the emission corona.}}
In: \beditor{\bsnm{{Ru{\v s}in, V., Heinzel, P., Vial, J.-C.}}} (ed.)
\bbtitle{Solar Coronal Structures, IAU Colloq. {\bf 144}},
\bfpage{487}.
\adsurl{1994scs..conf..487R}.
\end{bchapter}
\endbibitem

\bibitem[\protect\citeauthoryear{{Sakurai}
  \textit{et~al.}}{2002}]{2002SoPh..209..265S}
\begin{barticle}
\bauthor{\bsnm{{Sakurai}}, \binits{T.}},
\bauthor{\bsnm{{Ichimoto}}, \binits{K.}},
\bauthor{\bsnm{{Raju}}, \binits{K.P.}},
\bauthor{\bsnm{{Singh}}, \binits{J.}}:
\byear{2002},
\batitle{{Spectroscopic Observation of Coronal Waves}}.
\bjtitle{\solphys}
\bvolume{209},
\bfpage{265}.
\doiurl{10.1023/A:1021297313448}.
\adsurl{2002SoPh..209..265S}.
\end{barticle}
\endbibitem

\bibitem[\protect\citeauthoryear{{Singh}
  \textit{et~al.}}{1997}]{1997SoPh..170..235S}
\begin{barticle}
\bauthor{\bsnm{{Singh}}, \binits{J.}},
\bauthor{\bsnm{{Cowsik}}, \binits{R.}},
\bauthor{\bsnm{{Raveendran}}, \binits{A.V.}},
\bauthor{\bsnm{{Bagare}}, \binits{S.P.}},
\bauthor{\bsnm{{Saxena}}, \binits{A.K.}},
\bauthor{\bsnm{{Sundararaman}}, \binits{K.}},
\bauthor{\bsnm{{Krishan}}, \binits{V.}},
\bauthor{\bsnm{{Naidu}}, \binits{N.}},
\bauthor{\bsnm{{Samson}}, \binits{J.P.A.}},
\bauthor{\bsnm{{Gabriel}}, \binits{F.}}:
\byear{1997},
\batitle{{Detection of Short-Period Coronal Oscillations during the Total Solar
  Eclipse of 24 October, 1995}}.
\bjtitle{\solphys}
\bvolume{170},
\bfpage{235}.
\doiurl{10.1023/A:1004943924584}.
\adsurl{1997SoPh..170..235S}.
\end{barticle}
\endbibitem

\bibitem[\protect\citeauthoryear{{Singh}
  \textit{et~al.}}{2009}]{2009SoPh..260..125S}
\begin{barticle}
\bauthor{\bsnm{{Singh}}, \binits{J.}},
\bauthor{\bsnm{{Hasan}}, \binits{S.S.}},
\bauthor{\bsnm{{Gupta}}, \binits{G.R.}},
\bauthor{\bsnm{{Banerjee}}, \binits{D.}},
\bauthor{\bsnm{{Muneer}}, \binits{S.}},
\bauthor{\bsnm{{Raju}}, \binits{K.P.}},
\bauthor{\bsnm{{Bagare}}, \binits{S.P.}},
\bauthor{\bsnm{{Srinivasan}}, \binits{R.}}:
\byear{2009},
\batitle{{Intensity Oscillation in the Corona as Observed during the Total
  Solar Eclipse of 29 March 2006}}.
\bjtitle{\solphys}
\bvolume{260},
\bfpage{125}.
\doiurl{10.1007/s11207-009-9434-6}.
\adsurl{2009SoPh..260..125S}.
\end{barticle}
\endbibitem

\bibitem[\protect\citeauthoryear{{Singh}
  \textit{et~al.}}{2011}]{2011SoPh..270..213S}
\begin{barticle}
\bauthor{\bsnm{{Singh}}, \binits{J.}},
\bauthor{\bsnm{{Hasan}}, \binits{S.S.}},
\bauthor{\bsnm{{Gupta}}, \binits{G.R.}},
\bauthor{\bsnm{{Nagaraju}}, \binits{K.}},
\bauthor{\bsnm{{Banerjee}}, \binits{D.}}:
\byear{2011},
\batitle{{Spectroscopic Observation of Oscillations in the Corona During the
  Total Solar Eclipse of 22 July 2009}}.
\bjtitle{\solphys}
\bvolume{270},
\bfpage{213}.
\doiurl{10.1007/s11207-011-9732-7}.
\adsurl{2011SoPh..270..213S}.
\end{barticle}
\endbibitem

\bibitem[\protect\citeauthoryear{{Taroyan} and
  {Erd{\'e}lyi}}{2009}]{2009SSRv..149..229T}
\begin{barticle}
\bauthor{\bsnm{{Taroyan}}, \binits{Y.}},
\bauthor{\bsnm{{Erd{\'e}lyi}}, \binits{R.}}:
\byear{2009},
\batitle{{Heating Diagnostics with MHD Waves}}.
\bjtitle{\ssr}
\bvolume{149},
\bfpage{229}.
\doiurl{10.1007/s11214-009-9506-9}.
\adsurl{2009SSRv..149..229T}.
\end{barticle}
\endbibitem

\bibitem[\protect\citeauthoryear{{Tian}, {McIntosh}, and {De
  Pontieu}}{2011}]{2011ApJ...727L..37T}
\begin{barticle}
\bauthor{\bsnm{{Tian}}, \binits{H.}},
\bauthor{\bsnm{{McIntosh}}, \binits{S.W.}},
\bauthor{\bsnm{{De Pontieu}}, \binits{B.}}:
\byear{2011},
\batitle{{The Spectroscopic Signature of Quasi-periodic Upflows in Active
  Region Timeseries}}.
\bjtitle{\apjl}
\bvolume{727},
\bfpage{L37}.
\doiurl{10.1088/2041-8205/727/2/L37}.
\adsurl{2011ApJ...727L..37T}.
\end{barticle}
\endbibitem

\bibitem[\protect\citeauthoryear{{Tian}
  \textit{et~al.}}{2012}]{2012ApJ...759..144T}
\begin{barticle}
\bauthor{\bsnm{{Tian}}, \binits{H.}},
\bauthor{\bsnm{{McIntosh}}, \binits{S.W.}},
\bauthor{\bsnm{{Wang}}, \binits{T.}},
\bauthor{\bsnm{{Ofman}}, \binits{L.}},
\bauthor{\bsnm{{De Pontieu}}, \binits{B.}},
\bauthor{\bsnm{{Innes}}, \binits{D.E.}},
\bauthor{\bsnm{{Peter}}, \binits{H.}}:
\byear{2012},
\batitle{{Persistent Doppler Shift Oscillations Observed with Hinode/EIS in the
  Solar Corona: Spectroscopic Signatures of Alfv{\'e}nic Waves and Recurring
  Upflows}}.
\bjtitle{\apj}
\bvolume{759},
\bfpage{144}.
\doiurl{10.1088/0004-637X/759/2/144}.
\adsurl{2012ApJ...759..144T}.
\end{barticle}
\endbibitem

\bibitem[\protect\citeauthoryear{{Torrence} and
  {Compo}}{1998}]{1998BAMS...79...61T}
\begin{barticle}
\bauthor{\bsnm{{Torrence}}, \binits{C.}},
\bauthor{\bsnm{{Compo}}, \binits{G.P.}}:
\byear{1998},
\batitle{{A Practical Guide to Wavelet Analysis.}}
\bjtitle{Bull. Am. Met. Soc.}
\bvolume{79},
\bfpage{61}.
\doiurl{10.1175/1520-0477(1998)079<0061:APGTWA>2.0.CO;2}.
\adsurl{1998BAMS...79...61T}.
\end{barticle}
\endbibitem

\bibitem[\protect\citeauthoryear{{Tsubaki}}{1977}]{1977SoPh...51..121T}
\begin{barticle}
\bauthor{\bsnm{{Tsubaki}}, \binits{T.}}:
\byear{1977},
\batitle{{Periodic oscillations found in coronal velocity fields}}.
\bjtitle{\solphys}
\bvolume{51},
\bfpage{121}.
\doiurl{10.1007/BF00240450}.
\adsurl{1977SoPh...51..121T}.
\end{barticle}
\endbibitem

\bibitem[\protect\citeauthoryear{{Van Doorsselaere}, {Nakariakov}, and
  {Verwichte}}{2008}]{2008ApJ...676L..73V}
\begin{barticle}
\bauthor{\bsnm{{Van Doorsselaere}}, \binits{T.}},
\bauthor{\bsnm{{Nakariakov}}, \binits{V.M.}},
\bauthor{\bsnm{{Verwichte}}, \binits{E.}}:
\byear{2008},
\batitle{{Detection of Waves in the Solar Corona: Kink or Alfv{\'e}n?}}
\bjtitle{\apjl}
\bvolume{676},
\bfpage{L73}.
\doiurl{10.1086/587029}.
\adsurl{2008ApJ...676L..73V}.
\end{barticle}
\endbibitem

\bibitem[\protect\citeauthoryear{{Verwichte}
  \textit{et~al.}}{2010}]{2010ApJ...724L.194V}
\begin{barticle}
\bauthor{\bsnm{{Verwichte}}, \binits{E.}},
\bauthor{\bsnm{{Marsh}}, \binits{M.}},
\bauthor{\bsnm{{Foullon}}, \binits{C.}},
\bauthor{\bsnm{{Van Doorsselaere}}, \binits{T.}},
\bauthor{\bsnm{{De Moortel}}, \binits{I.}},
\bauthor{\bsnm{{Hood}}, \binits{A.W.}},
\bauthor{\bsnm{{Nakariakov}}, \binits{V.M.}}:
\byear{2010},
\batitle{{Periodic Spectral Line Asymmetries in Solar Coronal Structures from
  Slow Magnetoacoustic Waves}}.
\bjtitle{\apjl}
\bvolume{724},
\bfpage{L194}.
\doiurl{10.1088/2041-8205/724/2/L194}.
\adsurl{http://cdsads.u-strasbg.fr/abs/2010ApJ...724L.194V}.
\end{barticle}
\endbibitem

\bibitem[\protect\citeauthoryear{{Verwichte}
  \textit{et~al.}}{2013}]{2013A&A...552A.138V}
\begin{barticle}
\bauthor{\bsnm{{Verwichte}}, \binits{E.}},
\bauthor{\bsnm{{Van Doorsselaere}}, \binits{T.}},
\bauthor{\bsnm{{White}}, \binits{R.S.}},
\bauthor{\bsnm{{Antolin}}, \binits{P.}}:
\byear{2013},
\batitle{{Statistical seismology of transverse waves in the solar corona}}.
\bjtitle{\aap}
\bvolume{552},
\bfpage{A138}.
\doiurl{10.1051/0004-6361/201220456}.
\adsurl{2013A\%26A...552A.138V}.
\end{barticle}
\endbibitem

\bibitem[\protect\citeauthoryear{{Voulgaris}
  \textit{et~al.}}{2012}]{2012SoPh..278..187V}
\begin{barticle}
\bauthor{\bsnm{{Voulgaris}}, \binits{A.G.}},
\bauthor{\bsnm{{Gaintatzis}}, \binits{P.S.}},
\bauthor{\bsnm{{Seiradakis}}, \binits{J.H.}},
\bauthor{\bsnm{{Pasachoff}}, \binits{J.M.}},
\bauthor{\bsnm{{Economou}}, \binits{T.E.}}:
\byear{2012},
\batitle{{Spectroscopic Coronal Observations During the Total Solar Eclipse of
  11 July 2010}}.
\bjtitle{\solphys}
\bvolume{278},
\bfpage{187}.
\doiurl{10.1007/s11207-012-9929-4}.
\adsurl{2012SoPh..278..187V}.
\end{barticle}
\endbibitem

\bibitem[\protect\citeauthoryear{{Wang} and
  {Solanki}}{2004}]{2004A&A...421L..33W}
\begin{barticle}
\bauthor{\bsnm{{Wang}}, \binits{T.J.}},
\bauthor{\bsnm{{Solanki}}, \binits{S.K.}}:
\byear{2004},
\batitle{{Vertical oscillations of a coronal loop observed by TRACE}}.
\bjtitle{\aap}
\bvolume{421},
\bfpage{L33}.
\doiurl{10.1051/0004-6361:20040186}.
\adsurl{2004A\%26A...421L..33W}.
\end{barticle}
\endbibitem

\bibitem[\protect\citeauthoryear{{Williams}
  \textit{et~al.}}{2001}]{2001MNRAS.326..428W}
\begin{barticle}
\bauthor{\bsnm{{Williams}}, \binits{D.R.}},
\bauthor{\bsnm{{Phillips}}, \binits{K.J.H.}},
\bauthor{\bsnm{{Rudawy}}, \binits{P.}},
\bauthor{\bsnm{{Mathioudakis}}, \binits{M.}},
\bauthor{\bsnm{{Gallagher}}, \binits{P.T.}},
\bauthor{\bsnm{{O'Shea}}, \binits{E.}},
\bauthor{\bsnm{{Keenan}}, \binits{F.P.}},
\bauthor{\bsnm{{Read}}, \binits{P.}},
\bauthor{\bsnm{{Rompolt}}, \binits{B.}}:
\byear{2001},
\batitle{{High-frequency oscillations in a solar active region coronal loop}}.
\bjtitle{\mnras}
\bvolume{326},
\bfpage{428}.
\doiurl{10.1046/j.1365-8711.2001.04491.x}.
\adsurl{2001MNRAS.326..428W}.
\end{barticle}
\endbibitem

\bibitem[\protect\citeauthoryear{{Williams}
  \textit{et~al.}}{2002}]{2002MNRAS.336..747W}
\begin{barticle}
\bauthor{\bsnm{{Williams}}, \binits{D.R.}},
\bauthor{\bsnm{{Mathioudakis}}, \binits{M.}},
\bauthor{\bsnm{{Gallagher}}, \binits{P.T.}},
\bauthor{\bsnm{{Phillips}}, \binits{K.J.H.}},
\bauthor{\bsnm{{McAteer}}, \binits{R.T.J.}},
\bauthor{\bsnm{{Keenan}}, \binits{F.P.}},
\bauthor{\bsnm{{Rudawy}}, \binits{P.}},
\bauthor{\bsnm{{Katsiyannis}}, \binits{A.C.}}:
\byear{2002},
\batitle{{An observational study of a magneto-acoustic wave in the solar
  corona}}.
\bjtitle{\mnras}
\bvolume{336},
\bfpage{747}.
\doiurl{10.1046/j.1365-8711.2002.05764.x}.
\adsurl{2002MNRAS.336..747W}.
\end{barticle}
\endbibitem

\end{thebibliography}

\end{article} 

\end{document}